\documentclass[sn-nature]{sn-jnl}

\usepackage{graphicx}%
\usepackage{multirow}%
\usepackage{amsmath,amssymb,amsfonts}%
\usepackage{amsthm}%
\usepackage{mathrsfs}%
\usepackage[title]{appendix}%
\usepackage{xcolor}%
\usepackage{textcomp}%
\usepackage{manyfoot}%
\usepackage{booktabs}%
\usepackage{algorithm}%
\usepackage{algorithmicx}%
\usepackage{algpseudocode}%
\usepackage{listings}%
\usepackage{lineno}

\unnumbered

\begin{document}

\title[IMBH TDE]{A Tidal Disruption Event from an Intermediate-mass Black Hole Revealed by Comprehensive Multi-wavelength Observations}


\author*[1,2]{\fnm{Jialai} \sur{Wang}}\email{jialaiwang@mail.ustc.edu.cn}
\author[1,2]{\fnm{Mengqiu} \sur{Huang}}
\author*[1,2]{\fnm{Yongquan} \sur{Xue}}\email{xuey@ustc.edu.cn}
\author*[1,2]{\fnm{Ning} \sur{Jiang}}\email{jnac@ustc.edu.cn}
\author[1,2]{\fnm{Shifeng} \sur{Huang}}
\author[1,2]{\fnm{Yibo} \sur{Wang}}
\author[1,2]{\fnm{Jiazheng} \sur{Zhu}}
\author[1,2]{\fnm{Shifu} \sur{Zhu}}
\affil[1]{\orgdiv{Department of Astronomy}, \orgname{University of Science and Technology of China}, \orgaddress{\city{Hefei} \postcode{230026}, \country{China}}}
\affil[2]{\orgdiv{School of Astronomy and Space Science}, \orgname{University of Science and Technology of China}, \orgaddress{\city{Hefei} \postcode{230026}, \country{China}}}

\author[3]{\fnm{Lixin} \sur{Dai}}
\affil[3]{\orgdiv{Department of Physics}, \orgname{The University of Hong Kong}, \orgaddress{\street{Pokfulam Road}, \city{Hong Kong}, \country{China}}}

\author[4,5,6]{\fnm{Chichuan} \sur{Jin}}
\affil[4]{\orgdiv{National Astronomical Observatories}, \orgname{Chinese Academy of Sciences}, \orgaddress{\city{Beijing} \postcode{100101}, \country{China}}}
\affil[5]{\orgdiv{School of Astronomy and Space Science}, \orgname{University of Chinese Academy of Sciences}, \orgaddress{\city{Beijing} \postcode{100049}, \country{China}}}
\affil[6]{\orgdiv{Institute for Frontier in Astronomy and Astrophysics}, \orgname{Beijing Normal University}, \orgaddress{\city{Beijing} \postcode{102206}, \country{China}}}

\author[7,8]{\fnm{Bin} \sur{Luo}}
\affil[7]{\orgdiv{School of Astronomy and Space Science}, \orgname{Nanjing University}, \orgaddress{\city{Nanjing} \postcode{210093}, \country{China}}}
\affil[8]{\orgdiv{Key Laboratory of Modern Astronomy and Astrophysics (Nanjing University)}, \orgname{Ministry of Education}, \orgaddress{\city{Nanjing} \postcode{210093}, \country{China}}}

\author[9]{\fnm{Xinwen} \sur{Shu}}
\affil[9]{\orgdiv{Department of Physics}, \orgname{Anhui Normal University}, \orgaddress{\city{Wuhu}, \state{Anhui} \postcode{241002}, \country{China}}}

\author[10]{\fnm{Mouyuan} \sur{Sun}}
\affil[10]{\orgdiv{Department of Astronomy}, \orgname{Xiamen University}, \orgaddress{\city{Xiamen}, \state{Fujian} \postcode{361005}, \country{China}}}

\author[1,2]{\fnm{Tinggui} \sur{Wang}}

\author[11]{\fnm{Fan} \sur{Zou}}
\affil[11]{\orgdiv{Department of Astronomy}, \orgname{University of Michigan}, \orgaddress{\street{1085 S University}, \city{Ann Arbor}, \state{MI} \postcode{48109}, \country{USA}}}








\abstract{Tidal disruption events (TDEs) occur when a star crosses the tidal radius of a black hole (BH) and is ripped apart, providing a powerful way to probe dormant BHs over a wide mass range. In this study, we present our late-time observations and comprehensive multi-wavelength analyses of AT 2018cqh, a TDE at the center of a dwarf galaxy that exhibited successive flares in the optical, X-ray, and radio bands. We discovered an unexpected high-state X-ray plateau phase following the peak until the present time. Along with its reported prolonged rise lasting at least 550 days, these unique characteristics are consistent with the scenario of a TDE caused by an intermediate-mass black hole (IMBH) with a mass of approximately $(1-6) \times 10^5$ solar masses. Furthermore, scaling relations derived from the host-galaxy properties indicated a similar BH mass in concert. This discovery highlights the invaluable role of TDEs in the search for elusive IMBHs.}




\maketitle

\section{Introduction}\label{intro}

Intermediate-mass black holes (IMBHs) are a population of black holes that possess masses between stellar-mass (about $10 \, \mathrm{M}_\odot$) and supermassive (approximately ${10}^{6}-{10}^{10} \, \mathrm{M}_\odot$) black holes. They are typically less luminous than supermassive black holes and are generally more distant than stellar-mass black holes, which makes them particularly elusive \cite{Greene+2020}. Tidal disruption events (TDEs) are believed to be effective in detecting these mysterious objects \cite{Greene+2020,Reines2022}.

TDEs show that black holes at the centers of galaxies can disrupt stars that come too close \cite{Rees1988}. When a star crosses the tidal-disruption radius, it might be torn apart, generate a luminous flare as the debris falls back, and form an accretion disk around the black hole. This event serves as a valuable tool for detecting dormant black holes, especially when standard direct measurements are challenging \cite{Gezari2021}.

Most TDEs identified so far have been found to be powered by supermassive black holes. Only a few cases provide tentative evidence supporting the existence of IMBH-driven TDEs. Some optically selected TDEs, for example, suggest black hole masses $\lesssim 10^6 \, \mathrm{M}_\odot$ \cite{Yao+2023}, as inferred using the $M_{\rm BH} - \sigma_\star$ scaling relation. One such instance is AT 2020neh \cite{Angus+2022}, which stands out due to its unusually rapidly varying optical light curve. Despite the valuable observational insights provided by optical emission from TDEs, the exact mechanism behind the radiation remains uncertain \cite{Roth+2020}, with reprocessing \cite{Guillochon+2014,Roth+2016} and debris stream collisions \cite{Piran+2015} as possible explanations.

In contrast, soft X-ray emission, which is generally considered to arise primarily from the accretion disk around the black hole, offers more direct and reliable physical diagnostics \cite{Saxton+2020}. As such, X-ray observations are especially valuable for constraining black-hole parameters. This makes X-ray observations a more robust tool for studying TDEs, particularly those likely associated with IMBHs, such as 3XMM J152130.7$+$074916 \cite{Lin+2015}, 3XMM J215022.4$-$055108 \cite{Lin+2018}, and SDSS J134244.4$+$053056.1 \cite{He+2021}, all of which were tentatively identified based on their X-ray features. More recently, the newly discovered TDE EP240222a \cite{Jin+2025} has been spectroscopically confirmed to be situated at the outskirts of a galaxy, with optical and X-ray observations supporting the presence of an IMBH as the culprit of this TDE. However, these events lack sufficient sampling of the rise-to-peak X-ray light curves, making it difficult to establish a comprehensive understanding of the properties and evolutionary behaviors of IMBH TDEs.

Before the launch of the Einstein Probe (EP; \cite{Yuan+2025}) mission, the extended ROentgen Survey with an Imaging Telescope Array (eROSITA; \cite{Predehl+2021}) aboard the Spectrum-Roentgen-Gamma (SRG; \cite{Sunyaev+2021}) observatory conducted several all-sky survey observations, providing high-quality X-ray data sampling that could potentially capture key stages in the evolution of TDEs, including those associated with IMBHs. One notable example is AT 2018cqh (RA $=$ ${2}^{\rm h}{33}^{\rm m}{46.930}^{\rm s}$, Dec $=$ ${-1}^{\circ}{1}^{\prime}{28.38}^{\prime\prime}$), which was first discovered by the \textit{Gaia} Alerts team on 16 June 2018 as a new unclassified optical transient associated with the galaxy SDSS J023346.93$-$010128.3 ($z_{\rm spec} = 0.0489$) \cite{Delgado+2018} and later by the Asteroid Terrestrial Impact Last Alert System (ATLAS; \cite{Tonry+2018}) on 27 July 2018. After the optical flare emerged, peaked, and then faded, an X-ray counterpart (SRGe J023346.8$-$010129) was identified on 21 January 2020 by SRG/eROSITA during the first eROSITA All-Sky Survey (eRASS1) \cite{Bykov+2024}. The X-ray counterpart exhibited a significant outburst characteristic of a TDE candidate, showing a tenfold increase in X-ray flux from eRASS1 to eRASS4. This was followed by a decline in flux between eRASS4 and eRASS5. A follow-up observation with the X-ray Telescope (XRT; \cite{Burrows+2005}) onboard the Neil Gehrels Swift Observatory (hereafter Swift) also confirmed this decreasing trend. Overall, the X-ray brightening of AT 2018cqh lasted for at least 550 days, proving it to be a distinctive X-ray transient with a clear long-duration rising feature, which might be associated with a low-mass black hole \cite{Bykov+2024,Zhang+2024}. A late-time radio flare further adds to its unique characteristics \cite{Zhang+2024}. Based on the optical light curve, which exhibits a typical TDE structure (see Supplementary Fig. \ref{optlc}) and an almost constant blue color (see Supplementary Fig. \ref{optcolor}), as well as very soft X-ray spectra and distinct radio behavior, the multi-wavelength observations of AT 2018cqh are inconsistent with other nuclear transients (e.g., Bowen fluorescence flares \cite{Trakhtenbrot+2019a} or flares from changing-look AGNs \cite{Trakhtenbrot+2019b}) and more strongly support the TDE interpretation. In the light of its intriguing evolution, this transient warrants further follow-up and multi-wavelength studies to explore its nature.

In this work, we present the results of our multi-wavelength follow-up observations, which disclose a newly identified sustained high-state X-ray plateau phase that has been persisting for over a year. This, along with comprehensive multi-wavelength analyses, strongly suggests the IMBH-TDE scenario.

\section{Results}\label{result}

\subsection{High-state X-ray Plateau}\label{plateau}

To explore the nature of AT 2018cqh, we integrated archival data with our proposed follow-up observations to comprehensively analyze its evolution as a well-sampled multi-wavelength transient. Figure \ref{figlc} illustrates the multi-wavelength light curves of this object. These light curves provide almost continuous coverage from the pre-outburst phase through the post-outburst period.

Our newly acquired observations from Swift/XRT, EP/FXT, and XMM/EPIC unexpectedly revealed that the unabsorbed flux in the 0.3--2.0 keV band has remained at the flux level of about $4 \times {10}^{-13}$ erg s$^{-1}$ cm$^{-2}$, with no further decline observed yet, as shown in Fig. \ref{figlc}b. At a distance of 217 Mpc, the average X-ray flux during the apparent plateau phase corresponds to a luminosity of $L_{\rm 0.3-2.0 \, keV}$ approximately $2.4 \times {10}^{42}$ erg s$^{-1}$, while the observed peak X-ray flux translates to a luminosity of $L_{\rm 0.3-2.0 \, keV}$ approximately $5.5 \times {10}^{42}$ erg s$^{-1}$ \cite{Bykov+2024}. This apparent plateau phase, now persisting for over 500 days, strongly indicates that the X-ray radiation process is stable, signifying a unique shift in its emission behavior.

A key feature of the plateau spectra is the excess in the harder X-ray band compared to the rising phase. In the rising phase, the combined eROSITA spectrum is particularly soft (with an effective power-law photon index of $\Gamma = 5.5 \pm 0.5$) \cite{Bykov+2024}. However, during the plateau phase, the X-ray spectra become relatively harder, with spectral indices $\Gamma$ ranging from 3 to 4, as shown in Fig. \ref{figxspec}. These spectra can be modeled using only a blackbody component (or a thermal disk component), showing a nominal temperature $kT$ around 100 eV (or $T_{\rm in}$ around 150 eV), together with a weaker residual power-law component. This excess could be attributed to a Comptonization component, signaling the formation of a disk corona \cite{Guolo+2024}.

To model these spectra, we used an empirical Comptonization model to upscatter the photons from the blackbody spectrum into a harder power-law component (see Methods, subsection X-ray Observations and Data Analysis). The derived blackbody temperature, $kT$ around 63 eV, remains consistent across the plateau phase, suggesting that the accretion disk has remained in a steady accretion state during this long-lasting period. This temperature also matches that obtained from the co-added eROSITA spectra during the rising phase \cite{Bykov+2024}, reinforcing the interpretation that the spectral shape of the thermal component has not experienced significant changes since the disk formation.

Although recent findings suggest that an optical/UV plateau phase may be a common feature in the late stages of many TDEs \cite{Mummery+2024}, X-ray plateau phases remain rare, with most events exhibiting a shorter or more transient X-ray emission phase. Some TDEs display late-time X-ray behaviors characterized by a decay trend \cite{Jonker+2020,Guolo+2024}, as illustrated in Supplementary Fig. \ref{xraylc}. A few events show marginal evidence of a low-state plateau-like phase, with flux dropping by more than 1 dex from the peak. However, this nominal ``plateau'' is often a result of state transitions and limited sampling rather than a true plateau. Only a handful of known TDEs display high-state (or peak) X-ray plateau phases. One of the most notable examples is 3XMM J150052.0$+$015452, which was discovered as a decade-long TDE with a prolonged high-state X-ray phase \cite{Lin+2017} and later interpreted as an IMBH TDE in subsequent analyses \cite{Lin+2022,Cao+2023}. Similarly, the recently discovered IMBH TDE, EP240222a, also exhibits a peak X-ray plateau \cite{Jin+2025}. Additionally, late-time FUV observations suggest that the light curves of TDEs from low-mass black holes ($< 10^{6.5} \, \mathrm{M}_\odot$) tend to flatten at late times \cite{vanVelzen+2019}. Due to the limited sampling of X-ray light curves in most TDEs, it is challenging to capture all the different phases and monitor their evolution. As a result, previously identified TDEs with peak plateaus missed the crucial phase leading up to the high-state plateau. AT 2018cqh, however, stands out as a unique case. It could represent a rare opportunity to capture a TDE with a long-duration high-state X-ray plateau, encompassing its entire lifecycle from the initial rise and peak, through the decay, and into the plateau phase.

The occurrence of such high-state X-ray plateaus raises questions regarding the mechanisms that govern their formation, and whether they truly represent a distinct phase in the TDE lifecycle or if they are artifacts of insufficient light-curve sampling during long-term decay. Our well-sampled observations provide clear evidence of a plateau phase with stable emission, consistent with the saturated luminosity seen in super-Eddington accreting black holes \cite{Wang+1999,Watarai+2000,Mineshige+2000}. When a solar-type star is disrupted by an IMBH, the return rate of stellar debris can exceed the Eddington limit, triggering a super-Eddington accretion phase that may last for years \cite{Rees1988,Ulmer1999}. In this context, the accretion disk is expected to be a slim disk, with thermally stable radiation pressure-dominated regions due to radial advection \cite{Shakura+1973,Abramowicz+1988}. Here, photon diffusion to the disk surface takes longer than the radial motion of the accreted gas, which traps photons within the flow, saturating the luminosity around the Eddington limit \cite{Wang+1999,Watarai+2000,Mineshige+2000,Krolik+2012}. This saturated luminosity has been observed in other saturated accreting systems \cite{Wang+2013,Du+2015}, and similar evidence is found in two IMBH TDEs, 3XMM J150052.0$+$015452 \cite{Lin+2022,Cao+2023} and EP240222a \cite{Jin+2025,Li+2025}. The long-duration, stable high-state plateau observed in AT 2018cqh is consistent with this scenario. As a result, the bolometric luminosity of AT 2018cqh is expected to be close to the Eddington luminosity, providing a robust estimate of the black-hole mass based on the observed luminosity. Indeed, during the plateau phase, the bolometric luminosity of AT 2018cqh is measured at $2.0 \times {10}^{43}$ erg s$^{-1}$ (see Methods, subsection X-ray Observations and Data Analysis), which implies a black-hole mass of $1.5 \times 10^{5} \, \mathrm{M}_\odot$.

\subsection{Prolonged X-ray Rise}\label{rise}

The X-ray outburst of AT 2018cqh consists of a rising phase lasting at least 550 days, with the initial flux being an order of magnitude lower than the observed peak flux. Remarkably, this well-sampled rise represents one of the longest X-ray rising phases ever observed in TDEs \cite{Bykov+2024}. The eROSITA spectra of AT 2018cqh demonstrate that during the X-ray brightening phase prior to the plateau, the X-ray emission consistently exhibited a spectral shape with a blackbody temperature of $63 \pm 7$ eV, with no significant temperature variations across the eRASS \cite{Bykov+2024}. These soft spectra indicate that the TDE’s rising phase should begin much earlier.

Given that soft X-ray emission has been detected in early times for nearly all optically-selected TDEs exhibiting late-time X-ray brightening \cite{Guolo+2024}, and considering the earlier optical flare, the evidence suggests that the X-ray rise phase of AT 2018cqh likely began during the optical outburst, or at least by the time the optical flare had subsided. This took place roughly three years before the decline in X-ray emission, making one of the longest estimated rise times for TDEs, with the IMBH TDE EP240222a being the only comparable instance. However, the optical flare might have occurred in EP240222a but was missed, possibly due to observational limitations or lower flux during the early stages of the event. Numerical simulations have indicated that stellar debris from TDEs experiences self-collisions due to relativistic precession, causing optical emission and leading to the eventual circularization into an accretion disk \cite{Shiokawa+2015,Bonnerot+2016}. The protracted evolution of AT 2018cqh can likely be attributed to the time required for the disk to form and circularize through self-collisions within the debris stream.

In cases where the black hole is less massive, the weaker apsidal precession results in stream intersections occurring farther from the black hole, leading to mild collisions, reduced energy loss, and a longer circularization timescale \cite{Dai+2015,Wong+2022}. Stars are more likely to be disrupted when following low-$\beta$ orbits, where the penetration parameter $\beta$ quantifies how deeply the star enters the black-hole gravitational well. The approximately 3-year rise timescale for AT 2018cqh aligns with the model-predicted disk circularization timescale of $T_{\rm circ}$ about 3 years for a Sun-like star being tidally disrupted by an IMBH with a mass around $(2-4) \times 10^{5} \, \mathrm{M}_\odot$, orbiting with a penetration parameter of $\beta$ about 1 \cite{Jin+2025,Dai+2015,Wong+2022}. Although this first-order estimation does not account for other effects like stream-disk collisions that could accelerate circularization and may have some degeneracy in the parameters, the general behavior observed is consistent with an IMBH origin.

\subsection{X-ray Spectral Features}\label{temperature}

By leveraging X-ray spectral features, we can gain more constraints on the black-hole mass. As stellar debris falls toward the black hole, it might form a hot, optically thick structure with a characteristic blackbody temperature $kT$, which depends on both the black-hole mass $M_{\rm BH}$ and the mass accretion rate $\dot{M}$ \cite{Ulmer1999}. This relationship is given by $kT$ approximately $40 \, ({\dot{M}}/{\dot{M}_{\rm Edd}})^{1/4} ({M_{\rm BH}}/{10^{6} \, \mathrm{M}_\odot})^{-1/4} \,\, \text{eV}$, where $\dot{M}_{\rm Edd}$ is the Eddington accretion rate \cite{Saxton+2020}. Assuming accretion occurs at the Eddington limit $\dot{M} = \dot{M}_{\rm Edd}$ and using the observed blackbody temperature $kT$ of about 63 eV from eROSITA \cite{Bykov+2024}, we estimate the black-hole mass to be approximately $1.6 \times 10^5 \, \mathrm{M}_\odot$, which is consistent with aforementioned results.

However, the above single-temperature blackbody model might be unphysical, which could result in an underestimation of the accretion disk sizes \cite{Mummery+2021}. Therefore, we can consider the inner-disk temperature $T_{\rm in}$, which also depends on both the mass accretion rate $\dot{M}$ and the black-hole mass $M_{\rm BH}$, according to the standard thin-disk model \cite{Shakura+1973}. For a Schwarzschild black hole ($a = 0$), the inner-disk temperature is: $T_{\rm in} \propto 230 \, ({\dot{M}}/{\dot{M}_{\rm Edd}})^{1/4} ({M_{\rm BH}}/{10^{4} \, \mathrm{M}_\odot})^{-1/4} \,\, \text{eV}$ \cite{Jin+2025}. Due to the absence of fitting results for inner-disk temperature from eROSITA for non-hardening spectra, and to minimize the impact of potential variability during the plateau phase, we adopt the average temperature as a representative value. Using the average nominal inner-disk temperature $T_{\rm in}$ of about 153 eV and assuming Eddington-limited accretion, the estimated black-hole mass is approximately $5.1 \times \, 10^4 \, \mathrm{M}_\odot$. Alternatively, with the average decomposed inner-disk temperature $T_{\rm in}$ of about 82 eV, the mass estimate increases to approximately $6.2 \times 10^5 \, \mathrm{M}_\odot$.

The estimated black-hole masses supports the IMBH scenario, though real conditions could be more intricate. Variations in accretion disk geometry and radiation processes may cause the spectrum to deviate from models, affecting the X-ray shape \cite{Shakura+1973,Abramowicz+1988,Koratkar+1999,Davis+2005}. Black hole spin and general relativity effects can further complicate spectral analysis \cite{Novikov+1973,Done+2012,Wen+2022}. For the X-ray spectra of AT 2018cqh during the plateau phase, the spectral quality is not sufficient to properly constrain the weak power-law component, which could result in an overestimation of the power-law's contribution to the soft X-ray emission, potentially leading to a lower decomposed temperature and a higher black hole mass estimate \cite{Lin+2017}.

\subsection{Host Properties}\label{host}

\begin{figure}[p]
\centering
\includegraphics[width=0.9\textwidth]{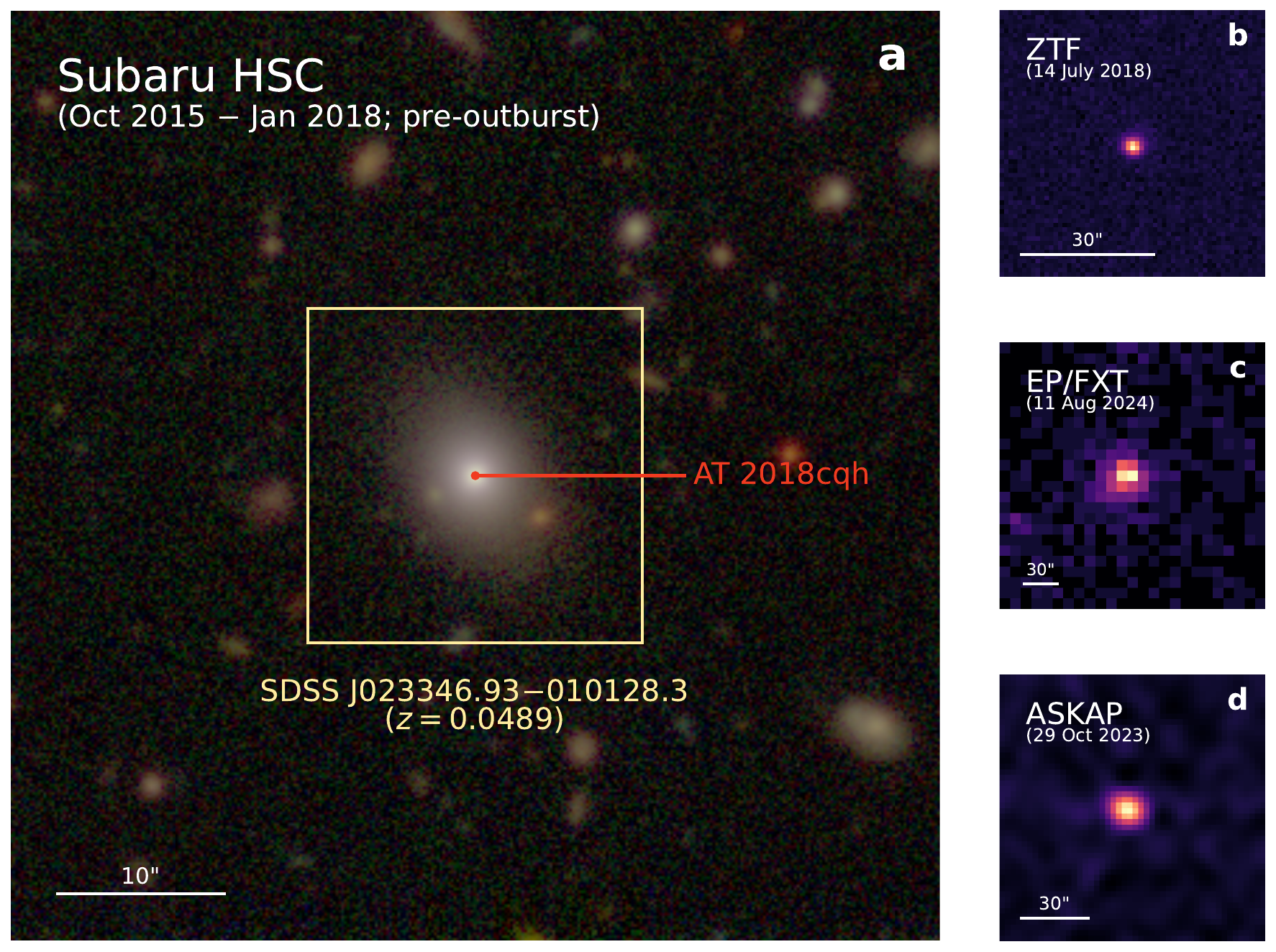}
\caption{\textbf{{\textbar} Pre-outburst optical image of AT 2018cqh alongside multi-wavelength observations during its outburst.} The optical flare was first detected on 16 June 2018, the X-ray outburst on 21 January 2020, and the radio flare on 25 June 2021. \textbf{a} pseudo-colored pre-outburst image captured by the Hyper Suprime-Cam (HSC) on the 8.2 m Subaru Telescope, utilizing $g$, $r$, and $z$ bands. The yellow square highlights the host galaxy SDSS J023346.93$-$010128.3, while the red dot indicates the location of the outburst. From \textbf{b} to \textbf{d}, the images of AT 2018cqh are presented, with observation dates indicated: ZTF observation in the $g$ band, EP/FXT observation in the 0.3--10 keV band, and ASKAP observation at 887.5 MHz.}\label{figimg}
\end{figure}

Including our proposed observations, AT 2018cqh has been observed across multiple wavelengths, both before and after its outburst. Figure \ref{figimg} shows a deep optical image of AT 2018cqh prior to the outburst along with multi-wavelength images captured during the event. Optical observations during the flare from the \textit{Gaia} mission pinpointed the precise location of AT 2018cqh at the center of a galaxy SDSS J023346.93$-$010128.3, with an offset of 59.3 mas from the optical centroid. Accounting for a 55 mas average offset between the \textit{Gaia} alert coordinates and the \textit{Gaia} DR2 \cite{Hodgkin+2021}, AT 2018cqh is located at a projected distance of $56 \pm 52.6$ pc from the galaxy's optical center. Multi-wavelength confirmation from X-ray and radio observations also supports the findings. This close proximity to the galactic center is most straightforwardly explained by an outburst originated from the nucleus, powered by the central black hole, rather than from an off-nuclear black hole.

\begin{figure}[p]
\centering
\includegraphics[width=0.9\textwidth]{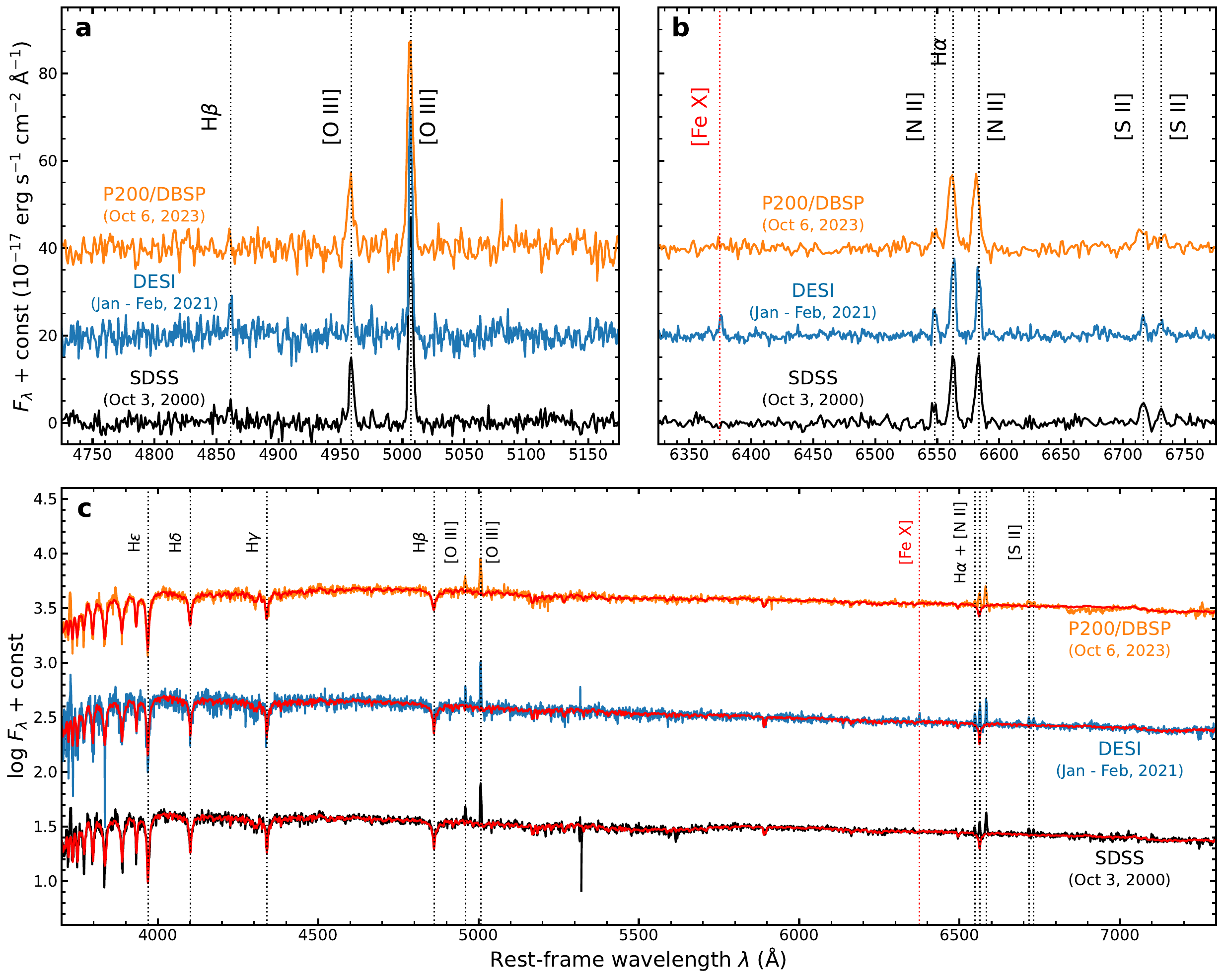}
\caption{\textbf{{\textbar} Multi-epoch optical spectra of AT 2018cqh and its host galaxy.} \textbf{a--b} the model-subtracted spectra at selected wavelengths, highlighting key features for detailed inspection. \textbf{c} The un-subtracted spectra along with the best-fit stellar models. The black spectrum in \textbf{a}, \textbf{b}, and \textbf{c} shows the spectrum recorded by SDSS on 3 October 2000, prior to the outburst. The blue and orange spectra present observations obtained during the X-ray outburst: the blue spectrum from DESI observations in early 2021 and the orange spectrum from the P200/DBSP observation on 6 October 2023. Overlaid red models in \textbf{c} are the best-fit stellar models for each spectrum. Vertical dotted lines highlight significant spectral features: the red vertical lines mark the transient coronal line [Fe \textsc{x}] $\lambda$6374, and the black vertical lines indicate other key lines, including the Balmer series (${\rm H}\alpha$, ${\rm H}\beta$, ${\rm H}\gamma$, ${\rm H}\delta$, and ${\rm H}\varepsilon$) and forbidden lines ([O \textsc{iii}], [N \textsc{ii}], and [S \textsc{ii}]). Source data are provided as a Source Data file.}\label{figoptspec}
\end{figure}

We also found that spectroscopic observations conducted with the Dark Energy Spectroscopic Instrument (DESI) during the X-ray outburst revealed a strong transient coronal line ([Fe \textsc{x}]$\lambda$6374) at the same redshift with a flux of $(1.4 \pm 0.2) \times 10^{-16}$ erg s$^{-1}$ cm$^{-2}$, which was absent in the pre-outburst spectrum from the Sloan Digital Sky Survey (SDSS), as shown in Fig. \ref{figoptspec}. Additionally, our observation using the Double Spectrograph (DBSP) on the Palomar 200-inch Hale Telescope (P200) during the X-ray plateau phase yielded an upper limit for the [Fe \textsc{x}]$\lambda$6374 line flux of $1.6 \times 10^{-16}$ erg s$^{-1}$ cm$^{-2}$ (at 90\% confidence). The presence of such transient coronal lines could be a unique diagnostic feature of TDEs \cite{Brandt+1995,Wang+2012}. This finding provides important evidence for the nature of the outburst and further strengthens that AT 2018cqh originated from the nucleus of the galaxy. Although AT 2018cqh exhibits a transient coronal line, its flux is insufficient to classify it as an extreme coronal line emitter (ECLE; \cite{Wang+2012,Hinkle+2024}). The host galaxy shows WISE and NEOWISE colors typical of non-active galaxies. The absence of a dust echo further indicates no substantial dust in the the galaxy's nucleus. This non-detection can be explained by the low metallicity of the host dwarf galaxy, which could contribute to the lack of a significant dust reservoir. It is also possible that the dust was sublimated during previous TDEs and has not yet reformed, especially in the case of a high TDE rate.

Nonetheless, the presence of a transient coronal line suggests that the event could be a TDE in a gas-rich environment \cite{vanVelzen+2021b}, which aligns with the outflow–cloud interaction model of the late-time radio flare \cite{Zhang+2024,Yang+2025}. Recent observations from the Australian Square Kilometre Array Pathfinder (ASKAP) also revealed that, after a brief peak in radio flux, the flux dropped to a local minimum and then began a second brightening phase, with the radio flux increasing nearly fourfold from the minimum to the latest observation. This makes the source one of the most luminous non-jetted TDEs observed at lower frequencies, as shown in Supplementary Fig. \ref{radiolc}. This unusual behavior sheds light on the complex environment of the circumnuclear medium.

The host galaxy of AT 2018cqh exhibits characteristics (e.g. stellar mass, color, and spectral features; see Methods, subsection Optical Observations and Data Analysis and subsection SED Fitting of the Pre-outburst Host) typical of a post-starburst (E+A) galaxy, with signs of recent star formation but no ongoing activity. Its pre-outburst stellar mass ($\log M_\star/\mathrm{M}_\odot = 9.3 \pm 0.1$) and dust-corrected rest-frame optical color ($u - r = 1.6 \text{ mag}$) indicate that it lies in the green valley \cite{Schawinski+2014,Hammerstein+2021}. The galaxy's pre-outburst spectral features show a low equivalent width of the ${\rm H}\alpha$ emission line (${\rm EW}_{\rm H\alpha} \lesssim 3 \, \text{\AA}$), indicating a lack of current star formation on timescales of about 10 Myr, and a strong Balmer absorption signature, as seen in the Lick ${\rm H}\delta_{\rm A}$ index (${\rm H}\delta_{\rm A} > 4 \, \text{\AA}$), pointing to a population of A-type stars formed during a past starburst around 1 Gyr ago. These features suggest that the galaxy is in a post-starburst phase, where star formation has largely ceased, leaving behind a stellar population that reflects both its recent star-forming history and its current quiescent state. Positioned on the ${\rm EW}_{\rm H\alpha}$ vs. Lick ${\rm H}\delta_{\rm A}$ diagram \cite{French+2016} (see Supplementary Fig. \ref{fighost}a), the galaxy falls into the post-starburst category, highlighting its rarity relative to typical SDSS galaxies. Notably, post-starburst galaxies like this one are over-represented among the hosts of TDEs, suggesting that their unique stellar populations and dynamic environments may play a role in the occurrence of such events \cite{French+2016,French+2020}.

Using the Baldwin–Phillips–Terlevich (BPT) diagram \cite{Baldwin+1981} (shown in Supplementary Fig. \ref{fighost}b), the host galaxy is located in the narrow-line active galactic nucleus (AGN) region, which is typically associated with Seyfert 2 galaxies. Given the absence of broad-line emission and the lack of AGN signatures in the X-ray spectra, infrared colors, and archival radio observations, the galaxy could be a recently ``dead'' AGN (or turned-off AGN) that the galaxy once harbored an AGN that has faded over a relatively short time, with its accretion disk no longer producing the intense radiation typically associated with an active black hole. Due to the light travel time to the narrow-line regions, their current ionization still reflects the AGN activity that occurred not long ago. Additionally, the absence of a detectable mid-infrared dust echo of AT 2018cqh further supports this scenario, indicating that the Type II-like appearance is due to the AGN having just turned off, rather than being caused by the obscuration of the torus.

Using a comprehensive dataset covering ultraviolet to mid-infrared wavelengths, we constrain the spectral energy distribution (SED) of the host galaxy prior to the outburst, allowing for a reliable estimate of the stellar mass. The detailed analysis yields a stellar mass of $(2.2 \pm 0.6) \times 10^9 \, \mathrm{M}_\odot$ (see Methods, subsection SED Fitting of the Pre-outburst Host), which is confidently in line with the typical properties of dwarf galaxies. 
The discrepancy between our result and previous estimates probably results from the absence of rest-frame near-infrared data in earlier studies, which has been demonstrated to cause an overestimation of stellar mass \cite{Ilbert+2010}. From the $M_\star - M_{\rm BH}$ correlation \cite{Greene+2020}, the corresponding black-hole mass is $\log M_{\rm BH}/\mathrm{M}_\odot = 5.6_{-0.4}^{+0.5}$ with an intrinsic scatter of 0.8 dex, placing it within the IMBH domain. The pre-outburst optical spectrum gives a stellar velocity dispersion of $\sigma_\star = 54 \pm 10 \, \text{km s}^{-1}$ (see Methods, subsection Optical Observations and Data Analysis). Using the $M_{\rm BH} - \sigma_\star$ relation \cite{Greene+2020}, this yields an estimated black-hole mass of $\log M_{\rm BH}/\mathrm{M}_\odot = 5.7_{-0.6}^{+0.5}$ with an intrinsic scatter of 0.5 dex. These characteristics of the optical host before the outburst lend support to the presence of an IMBH.

\subsection{Implications for IMBH TDEs}

The comprehensive multi-wavelength observations of AT 2018cqh have unveiled several fascinating properties, bolstering the case for its central engine being an IMBH. These observational results also offer valuable insights into the formation of its accretion disk and the underlying physics of accretion.

To summarize, AT 2018cqh stands as a well-sampled IMBH TDE with multi-wavelength features located in the center of a dwarf galaxy. It features one of the longest sampled X-ray rise times recorded for TDEs, as well as a long-lasting late-time high-state X-ray plateau phase which is still ongoing. This event is an IMBH TDE with well-sampled multi-wavelength light curves across optical, X-ray, and radio bands to date, capturing the entire lifecycle of the event --- from the rise and peak, through to the decay and plateau phases. This wealth of multi-wavelength outburst data from AT 2018cqh offers a unique opportunity to model and constrain the rare population of IMBH TDEs. The observed characteristics of AT 2018cqh are consistent with those of another recently discovered IMBH TDE, EP240222a, providing further confirmation of the typical behaviors exhibited by these intriguing and rare events, thereby deepening our understanding of the characteristics of this enigmatic class of IMBH TDEs.

\clearpage

\begin{figure}[p]
\centering
\includegraphics[width=0.9\textwidth]{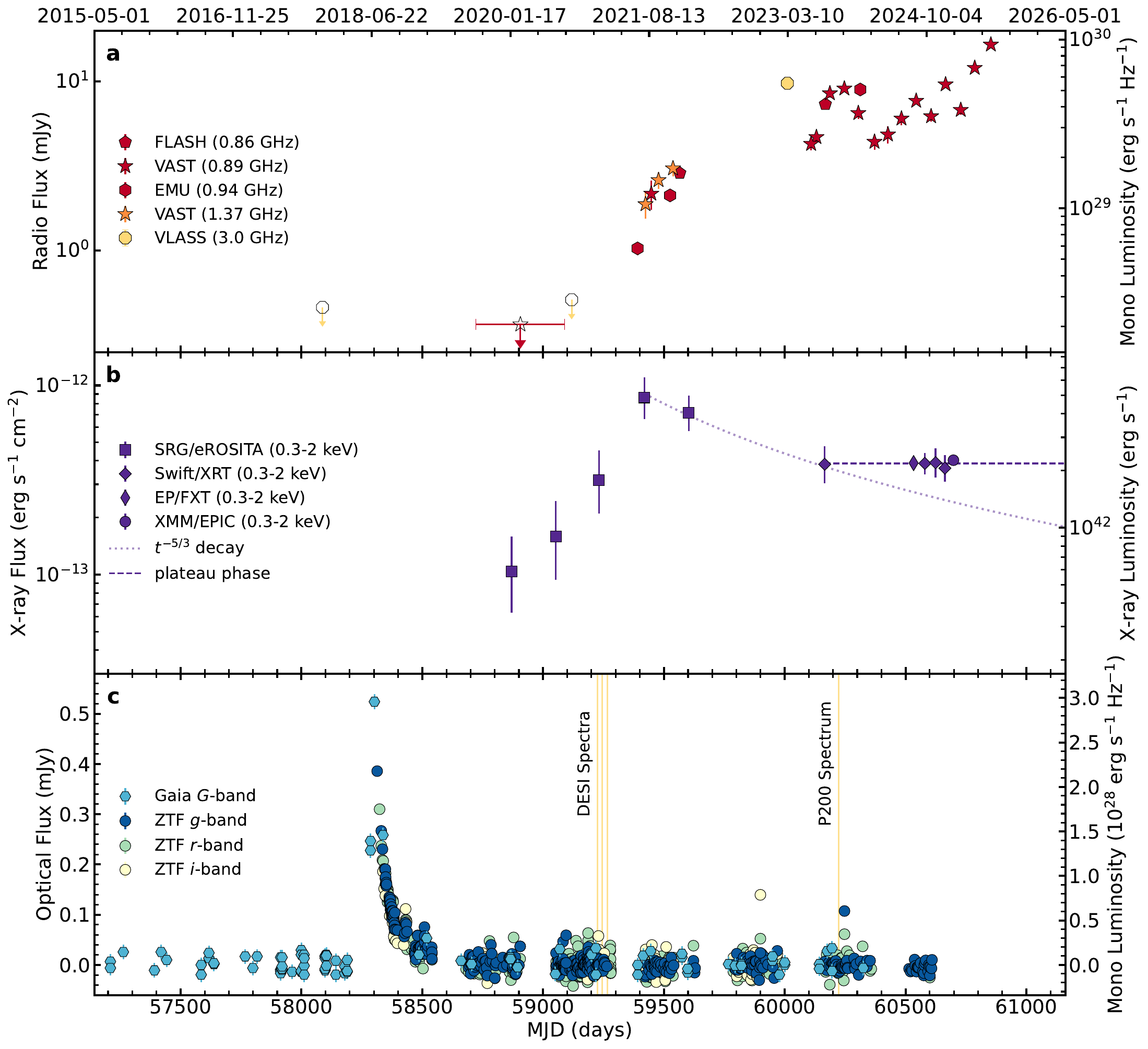}
\caption{\textbf{{\textbar} Multi-wavelength behavior of AT 2018cqh.} \textbf{a} radio light curves from the VLA (octagons) and ASKAP surveys (pentagons, stars, hexagons) across different frequencies. The colors transition from red to orange, indicating an increase in frequency. The downward arrows show the flux upper limits at 3$\sigma$ level. A comparison of the radio light curves from other events is presented in Supplementary Fig. \ref{radiolc}. \textbf{b} unabsorbed X-ray light curve in the 0.3--2.0 keV band, sourced from SRG/eROSITA (squares), Swift/XRT (diamonds), EP/FXT (thin diamonds), and XMM/EPIC (circles). All eROSITA data are taken from Fig. 10 in Bykov et al. (2024) \cite{Bykov+2024}. The violet dotted line represents a fit to the X-ray data collected after July 2021, based on the assumption that the flux follows a $t^{-5/3}$ decay trend. The dark violet dashed line indicates the plateau phase. The X-ray light curves of other optically-selected TDEs with late-time X-ray detections are compared in Supplementary Fig. \ref{xraylc}. \textbf{c} optical host-subtracted light curves in the \textit{Gaia} $G$-band (cyan hexagons), ZTF $g$-band (blue circles), ZTF $r$-band (green circles) and ZTF $i$-band (yellow circles). The gold vertical lines denote the observation periods for optical spectra obtained by DESI and P200/DBSP. The flare part of the optical light curves is zoomed in and shown in Supplementary Fig. \ref{optlc}, and the color evolution compared to other transient events is displayed in Supplementary Fig. \ref{optcolor}. The error bars represent 1$\sigma$ uncertainties. Source data are provided as a Source Data file.}\label{figlc}
\end{figure}

\clearpage

\begin{figure}[p]
\centering
\includegraphics[width=0.9\textwidth]{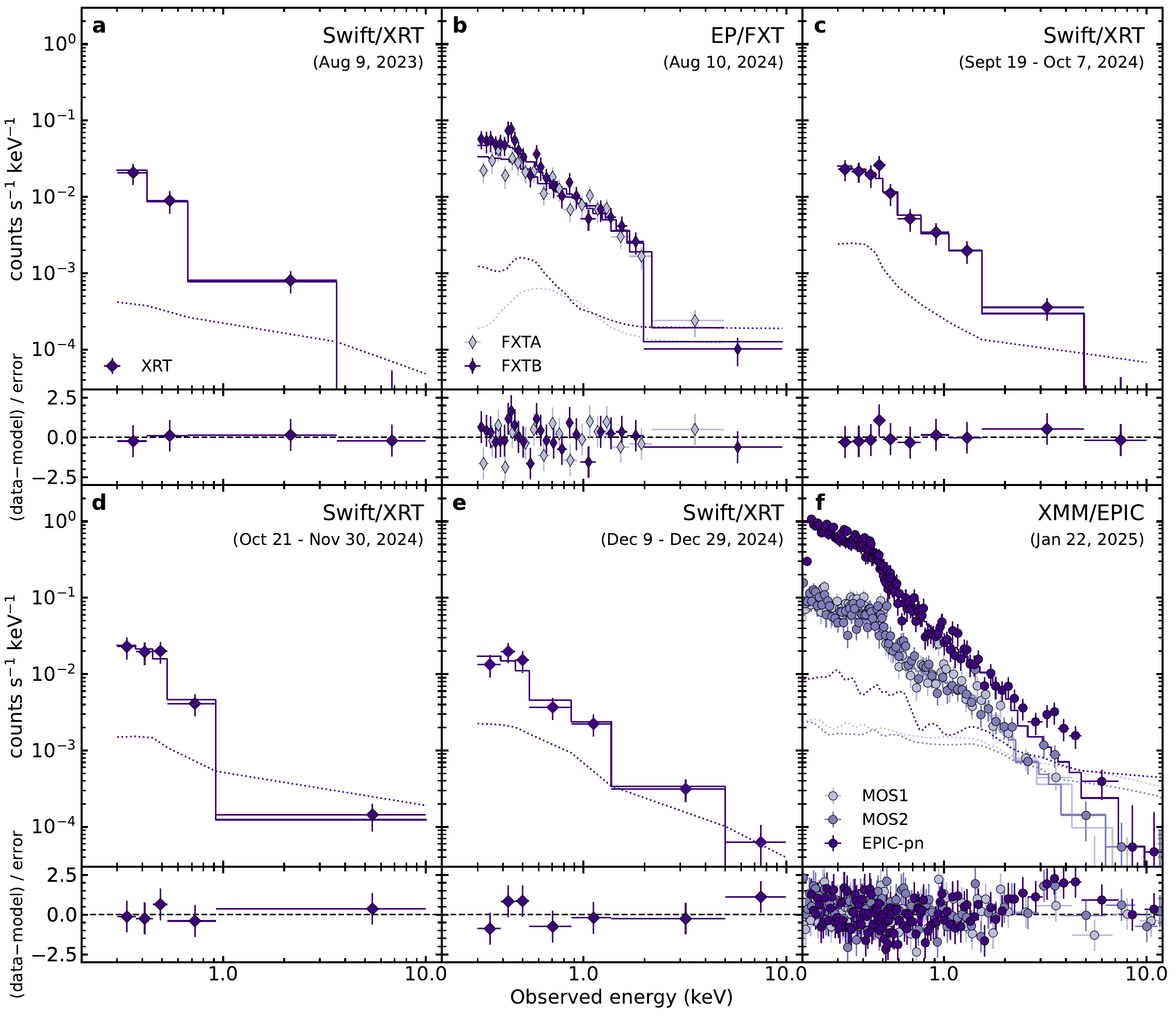}
\caption{\textbf{{\textbar} X-ray spectra of AT 2018cqh.} \textbf{a--f} includes the best-fit model shown as a solid line. Residuals for each spectrum are presented below to assess the fit quality. From \textbf{a} to \textbf{f}, the feature data includes Swift/XRT on 9 August 2023, EP/FXT on 10 August 2024, Swift/XRT from 19 September 2024 to 7 October 2024, Swift/XRT from 21 October 2024 to 30 November 2024, Swift/XRT from 9 December 2024 to 29 December 2024, and XMM/EPIC on 22 Jan 2025. For EP/FXT, the lighter spectrum represents FXTA, while the darker one corresponds to FXTB. For XMM/EPIC, the spectra are ordered from light to dark as follows: MOS1, MOS2, and EPIC-pn, which were fitted simultaneously. The dotted lines indicate the background level for each spectrum. The plotted data for each spectrum are rebinned to facilitate clearer display. The error bars represent 1$\sigma$ uncertainties. Source data are provided as a Source Data file.}\label{figxspec}
\end{figure}

\clearpage

\clearpage

\clearpage

\section{Methods}\label{method}

\subsection{X-ray Observations and Data Analysis}\label{xray}

The X-ray Telescope (XRT; \cite{Burrows+2005}) is an X-ray CCD imaging spectrometer aboard the Neil Gehrels Swift Observatory (Swift; \cite{Gehrels+2004}). AT 2018cqh was observed by follow-up target-of-opportunity (ToO) observations from August 2023 to December 2024. Swift/XRT data were processed using the HEASoft software suite \cite{Heasarc+2014}. We utilized the tasks \texttt{xrtpipeline}, \texttt{xselect}, and \texttt{xrtmkarf} to extract images, light curves, and spectra. The source and background spectra were extracted from circular regions centered on the source with a radius of 20 pixels, and from a nearby circular region with a radius of 50–60 pixels, respectively. To improve the signal-to-noise ratio (S/N), we combined observations from 19 September 2024 to 7 October 2024, 21 October 2024 to 30 November 2024, and 9 December 2024 to 29 December 2024, based on our approved ToO requests.
Swift/XRT continues to monitor AT 2018cqh based on our follow-up ToO requests at the time of writing of this paper.

The Follow-up X-ray Telescope (FXT; \cite{Chen+2020}) aboard the Einstein Probe (EP; \cite{Yuan+2025}) mission consists of two modules, each utilizing a Wolter-I telescope and pn-CCD detectors. After its launch on 9 January 2024, EP underwent a commissioning and calibration phase lasting approximately six months. The mission officially began scientific operations in July 2024. Our proposed EP Cycle-1 FXT Survey Target Observation (FSTO), conducted on 11 August 2024, involved an 11-ks exposure on AT 2018cqh. Data reduction was performed using Follow-up X-ray Telescope Data Analysis Software (FXTDAS)\footnote{https://epfxt.ihep.ac.cn/analysis} version 1.10, and we extracted scientific products including energy spectra, images, light curves, ancillary response files (ARF), redistribution matrix files (RMF), and background files, following the user guide v1.10. The same source and background regions used for Swift/XRT were applied in this process. For the analysis, the data from both FXT modules were modeled simultaneously.

The X-ray Multi-Mirror Mission (XMM-Newton; \citep{Jansen+2001}) carries two Reflection Grating Spectrometers (RGS), a European Photon Imaging Camera (EPIC) that includes a pn-CCD camera and two MOS imaging detectors, as well as an Optical Monitor (OM). We proposed a ToO observation with XMM-Newton, which was performed on 22 January 2025 with a total observation duration of 39 ks. The XMM-Newton data were reduced with the XMM-Newton Science Analysis System (SAS) \cite{Gabriel+2004} version 21.0.0. The \texttt{cifbuild} and \texttt{odfingest} tasks were executed for data preparation, followed by the extraction of light curves and spectra using the \texttt{xmmextractor} task. We modeled the EPIC-pn, MOS1, and MOS2 spectra simultaneously.

The X-ray observations used in this work are listed in Supplementary Table \ref{tabxlog}. Before fitting the spectra, all X-ray data were grouped to ensure at least 1 count per bin using grppha version 3.1.0. Spectral fitting was performed with XSPEC \cite{Arnaud+1996} version 12.15.0, using the Cash statistic in the 0.3--10 keV energy band for Swift and EP observations, and in the 0.2--12 keV energy band for the XMM-Newton observation. For each spectrum, three source models were applied: a power-law model ($powerlaw$), a blackbody model ($bbodyrad$), and a multi-color disk model ($diskbb$). To make a direct comparison with the results from Bykov et al. (2024) \cite{Bykov+2024} in Fig. \ref{figlc}, we used the same approach by omitting the redshift model ($zashift$). We then investigated the impact of including versus excluding the redshift model. Our test found no noticeable difference in flux and only a slight offset in temperature. Therefore, these models were only modified by Galactic absorption ($phabs$), resulting in the model combinations $phabs \times powerlaw$, $phabs \times bbodyrad$, and $phabs \times diskbb$. The fitting results are summarized in Supplementary Table \ref{tabxspec}.

For spectra obtained during the plateau phase, the photon indices are relatively harder, with a weak hard residual component compared to the spectra from the rise phase. Based on this, we adopted an empirical Comptonization model ($simpl$), where a fraction of the photons from a seed spectrum---either a blackbody ($bbodyrad$) or an accretion disk ($diskbb$)---undergoes upscattering to produce a harder power-law component. The models include Galactic absorption and are considered both with and without redshift, represented as $phabs \times (zashift) \times simpl \times bbodyrad$ and $phabs \times (zashift) \times simpl \times diskbb$. For FXT and EPIC, the photon indices of $simpl$ were treated as free parameters. Since the hard X-rays in the 2--10 keV range are not well constrained for the XRT spectra, we fixed $\Gamma = 2.3$ at the default value during the final modeling and presented both the free and fixed results. Figure \ref{figxspec} and Supplementary Table \ref{tabxspec} display the folded spectra and the corresponding fitting results.

We attempted to incorporate intrinsic absorption into the spectral fitting, but the results yielded abnormally small values, and no clear absorption-edge features were detected. This is likely due to the limited spectral quality, which may not be sufficient to resolve such features.

The bolometric luminosity during the plateau phase was derived by extrapolating the model fit ($phabs \times zashift \times simpl \times diskbb$) of the XMM/EPIC spectra to the 0.001--1000 keV range, assuming no Galactic absorption. We confirmed that emission outside this range is negligible, and thus, the derived luminosity could represent the full-band luminosity.

\subsection{Optical Observations and Data Analysis}\label{optical}

The \textit{Gaia} mission \cite{Gaia+2016} is an ESA astrometric mission designed to measure the three-dimensional spatial and velocity distributions of stars, as well as to determine their astrophysical properties. The \textit{Gaia} Photometric Science Alerts (``\textit{Gaia} Alerts'') is an all-sky photometric transient survey that uses repeated, high-precision measurements from the \textit{Gaia} satellite. The host galaxy of AT 2018cqh is covered by this survey, and the optical flare was captured in the process. We utilized the photometric data from the data processing pipeline for transient events. For the host-subtracted light curve, the reference flux was determined from the pre-outburst observations by calculating the median value.

The Zwicky Transient Facility (ZTF; \cite{Bellm+2019}) is an optical time-domain survey designed to conduct wide-area, high-cadence observations in search of rare astrophysical transients, variable stars, and moving objects. We obtained the ZTF light curve data from the ZTF Data Release 23 through the ZTF Science Data System \cite{Masci+2019}. Since the pre-outburst phase was not captured in the ZTF observations, we used the median flux value from the post-flare ZTF observations as the reference flux to generate the host-subtracted light curve for all bands.

The Sloan Digital Sky Survey (SDSS; \cite{York+2000}) has mapped over one-third of the sky, acquired deep multi-color images covering more than 8,000 square degrees, and obtained spectra for over three million astronomical objects. The host galaxy of AT 2018cqh was observed by the SDSS Legacy Survey on 3 October 2000, providing a reference spectrum for the pre-outburst phase, which we retrieved from SDSS Data Release 17 \cite{Abdurrouf+2022}.

The Dark Energy Spectroscopic Instrument (DESI; \cite{DESI+2022}) is an ongoing spectroscopic survey that aims to observe millions of galaxies, quasars, and stars. It will measure the position and recession velocity of approximately 40 million galaxies to construct detailed 3D maps of the Universe. The outburst spectra of AT 2018cqh were obtained on 10 January 2021, 30 January 2021, and 21 February 2021, during the Survey Validation phase of the DESI survey, which took place from December 2020 to May 2021, prior to the start of the DESI Main Survey. We obtained the co-added spectrum from the public Early Data Release \cite{DESI+2024}.

The Double Spectrograph (DBSP; \cite{Oke+1982}) on the Palomar 200-inch Hale Telescope (P200) is a low-to-medium resolution (R about 1,000 to 10,000) grating instrument. Our proposed observations on 6 October 2023 and 30 October 2024 captured the late-time spectra of AT 2018cqh during its plateau phase. We used the pypeit package to reduce the P200/DBSP spectra. Due to a low signal-to-noise (S/N) ratio, we excluded the observation from 30 October 2024 in our analysis.

We corrected all photometric data in the optical light curves for Galactic extinction using the Fitzpatrick (1999) extinction law \cite{Fitzpatrick+1999}, with $R_V = 3.1$ and a Galactic extinction of $E(B-V) = 0.0268 \pm 0.0003$ mag \cite{Schlafly+2011}.

To fit the stellar population in each spectrum, we employed the pPXF package \cite{Cappellari+2004} to derive the best-fit model. For the analysis of the host galaxy using the pre-outburst SDSS spectrum, we utilized measurements from the SDSS MPA-JHU catalogs \cite{Brinchmann+2004}, including the velocity dispersion for the $M_{\rm BH} - \sigma_\star$ relation, the fluxes of ${\rm H}\alpha$, ${\rm H}\beta$, [O \textsc{iii}], and [N \textsc{ii}] for BPT classification, as well as the absorption-corrected ${\rm H}\alpha$ equivalent widths and Lick ${\rm H}\delta$ indices for the identification of post-starburst galaxies.

For the scaling relations, we adopt those presented in Supplemental Table 5 of Greene et al. (2020) \cite{Greene+2020}, which were derived using data from all galaxies, including those with upper limits on $M_{\rm BH}$.

\subsection{Radio Observations}\label{radio}

The Australian Square Kilometre Array Pathfinder (ASKAP; \cite{Hotan+2021}) is a precursor telescope of the Square Kilometre Array (SKA), designed to observe mid-frequency radio waves with high survey speed. ASKAP is currently conducting several survey science projects aimed at investigating the structure and evolution of the Universe. AT 2018cqh lies within the fields covered by several ASKAP surveys, including the Rapid ASKAP Continuum Survey (RACS; \cite{McConnell+2020}), the Variables and Slow Transients Survey (VAST; \cite{Murphy+2021}), the Evolutionary Map of the Universe Pilot Survey (EMU; \cite{Norris+2021}), and the First Large Absorption Survey in H\textsc{i} (FLASH; \cite{Allison+2022}). Image data and source catalogs are sourced from the CSIRO ASKAP Science Data Archive (CASDA). The radio detection for AT 2018cqh was made using the Selavy sourcefinder, which identifies sources in an image and fits components to them.

The Very Large Array Sky Survey (VLASS; \cite{Lacy+2020}) is a synoptic, all-sky radio sky survey conducted using the Karl G. Jansky Very Large Array (VLA). The survey covers the entire sky visible to the VLA, specifically the region with a declination of $>-40^{\circ}$, which amounts to 33,885 square degrees. VLASS is designed to observe the sky in three distinct epochs, enabling the detection of variable and transient radio sources over time. AT 2018cqh has been observed across three VLASS epochs: VLASS 1.1 (2019), VLASS 2.1 (2021), and VLASS 3.1 (2023). For this study, we utilized the released VLASS Quick Look images and employed the \texttt{IMFIT} task in the Common Astronomy Software Applications (CASA) package \cite{CASA+2022} to extract relevant measurements.

\subsection{SED Fitting of the Pre-outburst Host}\label{sed}

To perform the SED fitting of the pre-outburst host galaxy, we compiled archival multi-wavelength data from different surveys. For the ultraviolet (UV) data, we used photometry in the far-ultraviolet (FUV) and near-ultraviolet (NUV) bands from the Galaxy Evolution Explorer (GALEX; \cite{Martin+2005}) mission. Optical data were sourced from SDSS for the u-band and from the Pan-STARRS1 surveys \cite{Chambers+2016} for the stacked $g$, $r$, $i$, $z$, and $y$ bands. In the near-infrared (NIR) range, we incorporated photometric measurements from the UKIRT Infrared Deep Sky Survey (UKIDSS; \cite{Lawrence+2007}) in the $Y$, $J$, $H$, and $K$ bands. For mid-infrared (MIR) data, we selected photometry from the Wide-field Infrared Survey Explorer (WISE; \citep{Wright+2010}) mission, spanning the $W1$, $W2$, $W3$, and $W4$ bands (3.4--22 $\mu\mathrm{m}$). We corrected all photometric data for Galactic extinction using the Fitzpatrick (1999) extinction law \cite{Fitzpatrick+1999}, with $R_V = 3.1$ and a Galactic extinction of $E(B-V) = 0.0268 \pm 0.0003$ mag \cite{Schlafly+2011}.

To analyze the SED and estimate the stellar mass of the pre-outburst host galaxy, we employed the Code Investigating GALaxy Emission (CIGALE; \cite{Boquien+2019,Yang+2020,Yang+2022}) v2022.1. We adopted a delayed star-formation history (SFH) model \cite{Boquien+2019} and utilized the BC03 stellar population templates \cite{Bruzual+2003}, with a Chabrier initial mass function (IMF) \cite{Chabrier+2003}. Nebular emission was included self-consistently through CLOUDY photoionization calculations \cite{Ferland+2017,Villa-Velez+2021}. For dust attenuation, we applied the flexible version of the Calzetti law \cite{Calzetti+2000}, while for the infrared dust emission, we used empirical templates \cite{Dale+2014} to model the galaxy's infrared emission.

Supplementary Fig. \ref{figsed} shows the best-fit SED. The best-fit stellar mass is estimated to be $1.9 \times 10^9 \, \mathrm{M}_\odot$, with a reduced chi-squared value of  $\chi^2_{\text{red}} = 0.45$. Additionally, CIGALE provided a Bayesian estimate of the stellar mass, yielding a value of $(2.2 \pm 0.6) \times 10^9 \, \mathrm{M}_\odot$. 

Based on the best-fit SED, we calculated the dust-corrected rest-frame $u - r$ color to be $1.6 \, \text{mag}$, representing the intrinsic color of the pre-outburst host galaxy.

We attempted to include an AGN contribution in the SED model, but the reduced chi-square of the best fit is always higher compared to the pure galaxy model, suggesting that the pre-outburst SED may not support the presence of an AGN.

\backmatter

\section{Data availability}

All data presented in this paper are publicly available. Swift data can be accessed via the HEASARC archives at https://heasarc.gsfc.nasa.gov/cgi-bin/W3Browse/w3browse.pl, EP data via the Einstein Probe Data Archive at https://ep.bao.ac.cn/ep/next/data-release, XMM data via the XMM-Newton Science Archive at https://nxsa.esac.esa.int, \textit{Gaia} data via the \textit{Gaia} Alerts at http://gsaweb.ast.cam.ac.uk/alerts/alert/Gaia18bod/, ZTF data via the IRSA at https://irsa.ipac.caltech.edu/Missions/ztf.html, ASKAP data via the CASDA at https://data.csiro.au/domain/casda, and VLASS Quick Look images via the NRAO website at https://vlass-dl.nrao.edu/vlass/quicklook. Source data are provided with this paper.

\section{Code availability}

Data reduction, X-ray spectral fitting, and SED fitting were conducted using standard, publicly available software packages, including HEASoft (https://heasarc.gsfc.nasa.gov/docs/software/lheasoft/), FXTDAS (https://epfxt.ihep.ac.cn/analysis), SAS (https://www.cosmos.esa.int/web/xmm-newton/sas), XSPEC (https://heasarc.gsfc.nasa.gov/docs/software/xspec/), and CIGALE (https://cigale.lam.fr/).


\bibliography{main}

\section{Acknowledgements}

We acknowledge W. N. Brandt for his suggestions. We acknowledge Marat Gilfanov, Sergei Bykov, and Rashid Sunyaev for their discussions. We acknowledge Weiyu Wu for obtaining the P200/DBSP spectrum on 30 October 2024. We acknowledge the Swift team and the XMM-Newton team for approving our ToO requests and carrying out the observations.
J.W., M.H., Y.X., and S.Z. acknowledge the support from the National Science Foundation of China grants (12393814 and 12025303), the Strategic Priority Research Program of the Chinese Academy of Sciences (grant No. XDB0550300), and the National Key R\&D Program of China (2023YFA1608100 and 2022YFF0503401).
N.J. acknowledges the support of the National Natural Science Foundation of China (12522303, 12192221) and the Strategic Priority Research Program of the Chinese Academy of Sciences (XDB0550200).
S.H. acknowledges the Anhui Provincial Natural Science Foundation (2308085QA32) and the Fundamental Research Funds for Central Universities (WK2030000097).
Y.W. acknowledges the National Natural Science Foundation of China grants (12503017), the Postdoctoral Fellowship Program of CPSF (Grant No. GZC2025209), the Fundamental Research Funds for Central Universities (WK2030250127), the China Postdoctoral Science Foundation (2025M773198), and the Anhui Provincial Natural Science Foundation (2508085QA006).
L.D. acknowledges the support from the National Natural Science Foundation of China and the Hong Kong Research Grants Council (N\_HKU782/23, 12122309, 17314822, 17304821).
C.J. acknowledges the National Natural Science Foundation of China through grant 12473016, and the support by the Strategic Priority Research Program of the Chinese Academy of Sciences (Grant No. XDB0550200).
This work is based on the data obtained with Einstein Probe, a space mission supported by Strategic Priority Program on Space Science of Chinese Academy of Sciences, in collaboration with ESA, MPE and CNES (Grant No. XDA15310000). 
We acknowledge the use of public data from the Swift data archive. 
The work is based on observations obtained with XMM-Newton, an ESA science mission with instruments and contributions directly funded by ESA Member States and NASA. 
This research uses data obtained through the Telescope Access Program (TAP). Observations with the Hale Telescope at Palomar Observatory were obtained as part of an agreement between the National Astronomical Observatories, the Chinese Academy of Sciences, and the California Institute of Technology. 
This paper includes archived data obtained through the CSIRO ASKAP Science Data Archive, CASDA (http://data.csiro.au). 

\section{Author contributions}

J.W. led the project and wrote the manuscript. Y.X. and N.J. coordinated the scientific investigations. M.H., S.H., and S.Z. contributed to the X-ray observations and data analysis. Y.W. and J.Z. assisted with the optical observations and data analysis. L.D. helped with the circularization timescale calculation. C.J., B.L., X.S., M.S., T.W., and F.Z. provided key suggestions. All authors discussed the results and commented on the manuscript.

\section{Competing interests}

The authors declare no competing interests.

\section{Supplementary Information}

\newcounter{Efigure}
\setcounter{Efigure}{0}
\renewcommand\figurename{Supplementary Fig.}
\renewcommand{\thefigure}{\arabic{Efigure}}
\setcounter{table}{0}

\begin{figure}[p]
\addtocounter{Efigure}{1}
\centering
\includegraphics[width=0.9\textwidth]{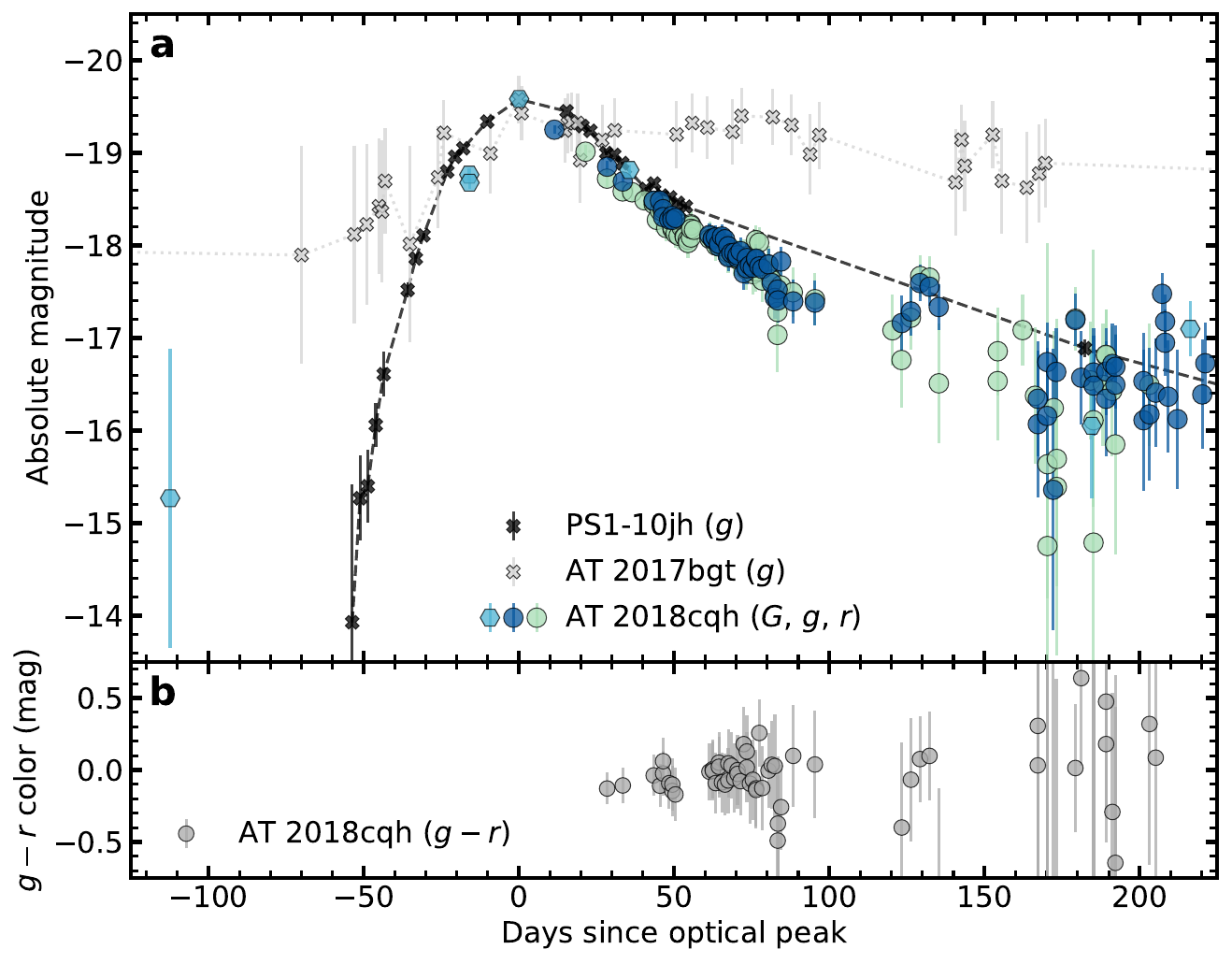}
\caption{\textbf{{\textbar} Optical behavior of AT 2018cqh. a} optical host-subtracted light curves in the \textit{Gaia} $G$-band (cyan hexagons), ZTF $g$-band (blue circles), ZTF $r$-band (green circles). For comparison, the black crosses and dashed black line represent the $g$-band light curve of the TDE PS1-10jh \cite{Gezari+2012}, while the gray crosses and dotted gray line show the $g$-band light curve of the Bowen fluorescence flare (BFF) AT 2017bgt \cite{Trakhtenbrot+2019a}. Both events are matched in peak absolute magnitude. \textbf{b} host-subtracted optical color evolution in the ZTF $g-r$ color, shown by gray circles. The error bars represent 1$\sigma$ uncertainties. Source data are provided as a Source Data file.}\label{optlc}
\end{figure}

\clearpage

\begin{figure}[p]
\addtocounter{Efigure}{1}
\centering
\includegraphics[width=0.9\textwidth]{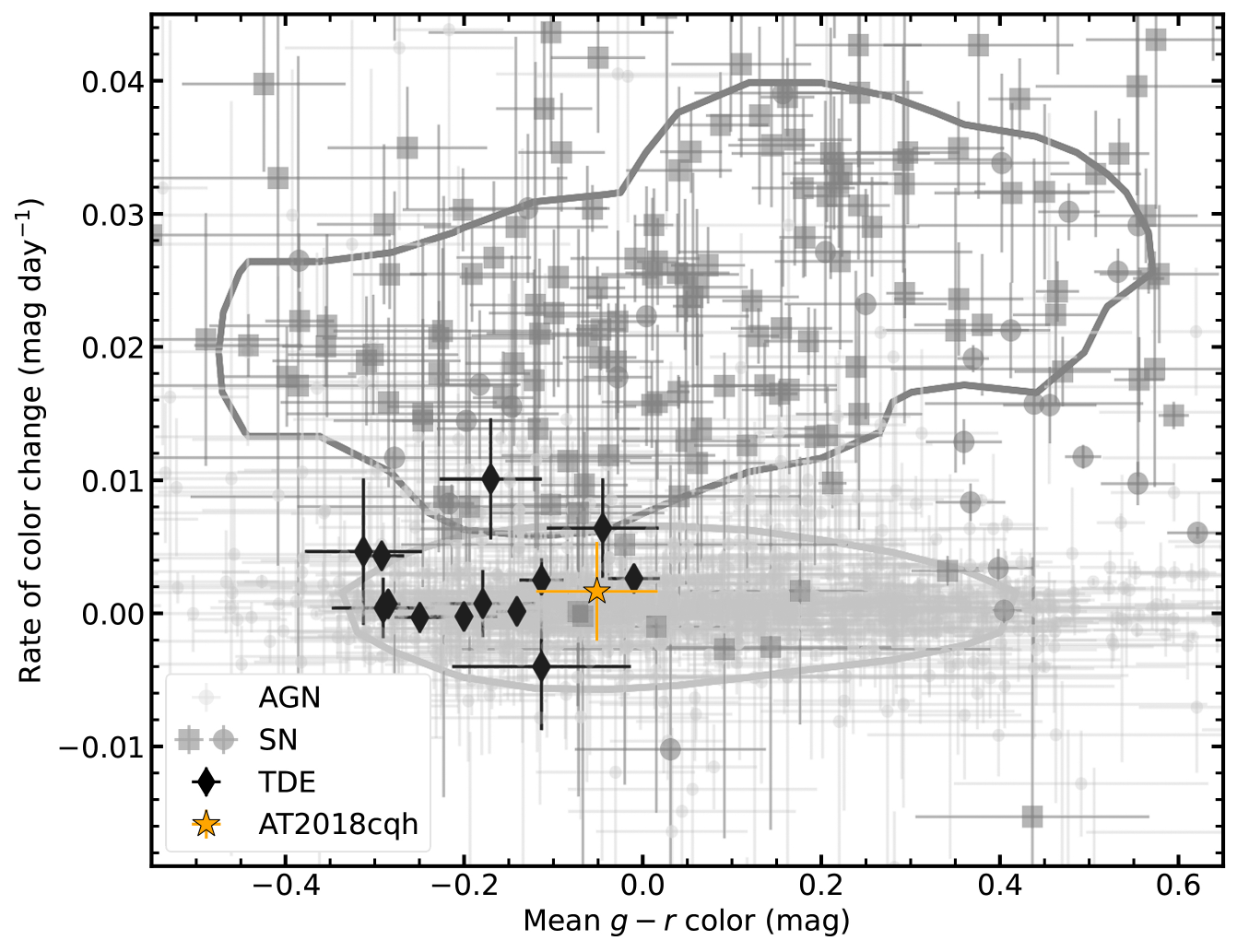}
\caption{\textbf{{\textbar} Mean $g-r$ color and color change ($\Delta(g-r)/t$) of AT 2018cqh.} The mean $g-r$ color is measured from all good detections in the ZTF light curve. The color change is calculated using a 3-day binned ZTF light curve to reduce the impact of false variations caused by detection uncertainties. The background is adapted from Fig. 1 in van Velzen et al. (2021a) \cite{vanVelzen+2021a}. All background events are nuclear transients from the first half of ZTF survey observations. The dark gray contours enclose two-thirds of the spectroscopically classified supernovae (dark gray circles and squares), while the light gray contours enclose two-thirds of the AGNs (light gray circles) in the sample \cite{vanVelzen+2021a}. AT 2018cqh (orange star) and other TDEs (black thin diamonds) show an almost constant optical color, whereas most supernovae show a color increase in post-peak observations. The error bars represent 1$\sigma$ uncertainties. Source data are provided as a Source Data file.}\label{optcolor}
\end{figure}

\clearpage

\begin{figure}[p]
\addtocounter{Efigure}{1}
\centering
\includegraphics[width=0.9\textwidth]{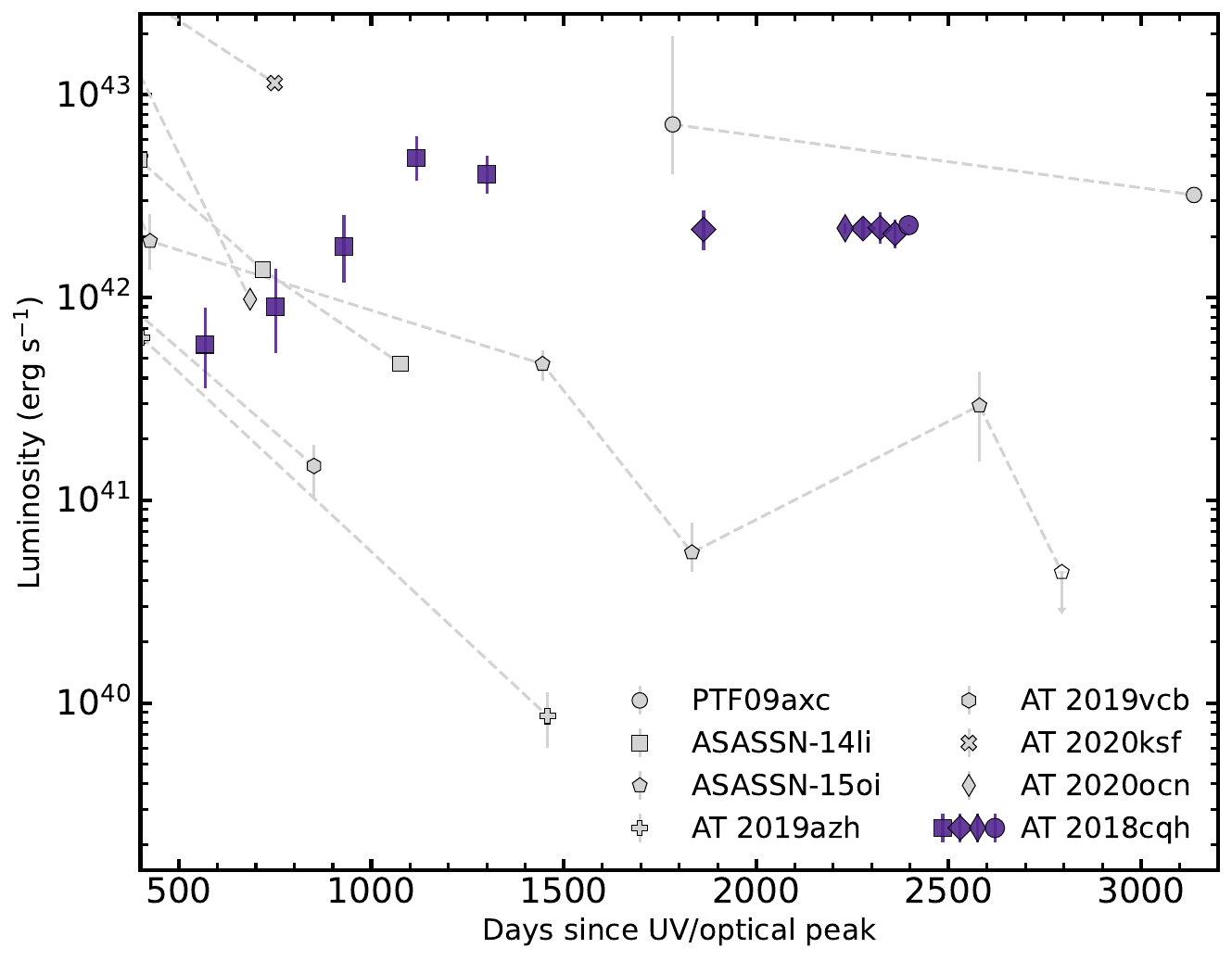}
\caption{\textbf{{\textbar} Unabsorbed X-ray luminosity light curve of AT 2018cqh.} The unabsorbed X-ray luminosity of AT 2018cqh is shown with magenta scatter points, sourced from SRG/eROSITA (violet squares), Swift/XRT (violet diamonds), EP/FXT (violet thin diamonds), and XMM/EPIC (violet circles). All eROSITA data are taken from Fig. 10 in Bykov et al. (2024) \cite{Bykov+2024}. For comparison, the X-ray luminosity light curves of seven optically-selected TDEs with late-time X-ray detections are also included: ASASSN-14li (0.3--10 keV; gray squares; \cite{Guolo+2024}), ASASSN-15oi (0.2--12 keV; gray pentagons; \cite{Hajela+2025}), AT 2019azh (0.3--10 keV; gray plus signs; \cite{Guolo+2024}), AT 2019vcb (0.3--10 keV; gray hexagons; \cite{Guolo+2024}), AT 2020ksf (0.3--10 keV; gray x signs; \cite{Guolo+2024}), AT 2019vcb (0.3--10 keV; gray thin diamonds; \cite{Guolo+2024}), and PTF09axc (0.3--7 keV; gray circles; \cite{Arcavi+2014,Jonker+2020}). The error bars represent 1$\sigma$ uncertainties. Source data are provided as a Source Data file.}\label{xraylc}
\end{figure}

\clearpage

\begin{figure}[p]
\addtocounter{Efigure}{1}
\centering
\includegraphics[width=0.9\textwidth]{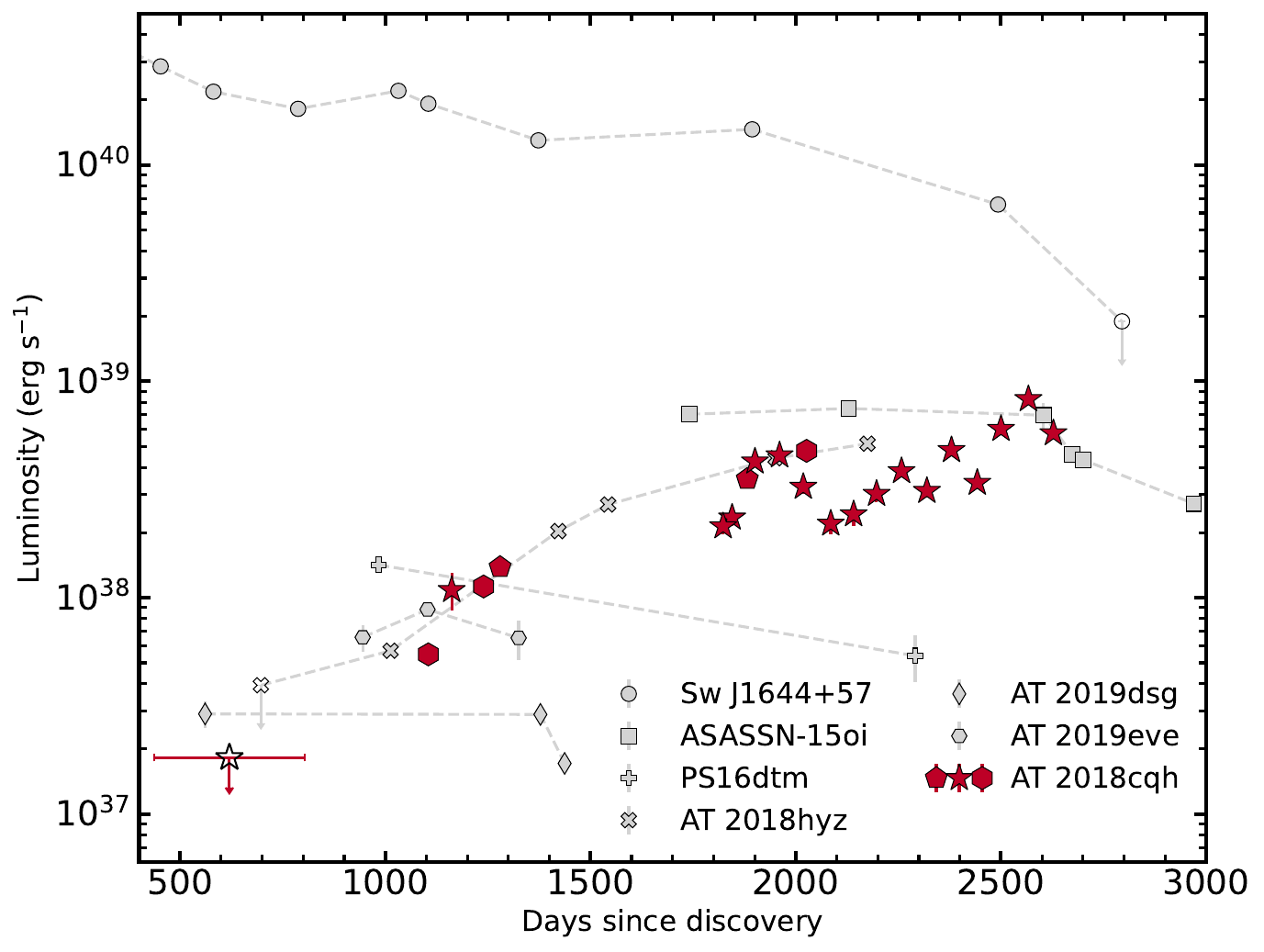}
\caption{\textbf{{\textbar} Radio luminosity light curves of AT 2018cqh.} The golden scatter points represent observations from ASKAP surveys: FLASH (red pentagons), VAST (red stars), and EMU (red hexagons) in 0.86--0.94 GHz. For comparison, radio luminosity light curves of other TDEs are shown, including the relativistic TDE Sw J1644$+$57 (1.4--1.8 GHz; gray circles; \cite{Berger+2012,Zauderer+2013,Eftekhari+2018,Cendes+2021b}), the non-relativistic event AT 2019dsg (1.4--1.5 GHz; gray thin diamonds; \cite{Cendes+2021a,Cendes+2024}), and four events with apparent late-rising radio emission: ASASSN-15oi (0.8--1.3 GHz; gray squares; \cite{Hajela+2025}), PS16dtm (0.9--1.7 GHz; gray plus signs; \cite{Cendes+2024}), AT 2018hyz (0.8--0.9 GHz; gray x signs; \cite{Cendes+2022,Cendes+2025}), and AT 2019eve (1.2--1.4 GHz; gray hexagons; \cite{Cendes+2024}). The error bars represent 1$\sigma$ uncertainties. Source data are provided as a Source Data file.}\label{radiolc}
\end{figure}

\clearpage

\begin{figure}[p]
\addtocounter{Efigure}{1}
\centering
\includegraphics[width=0.9\textwidth]{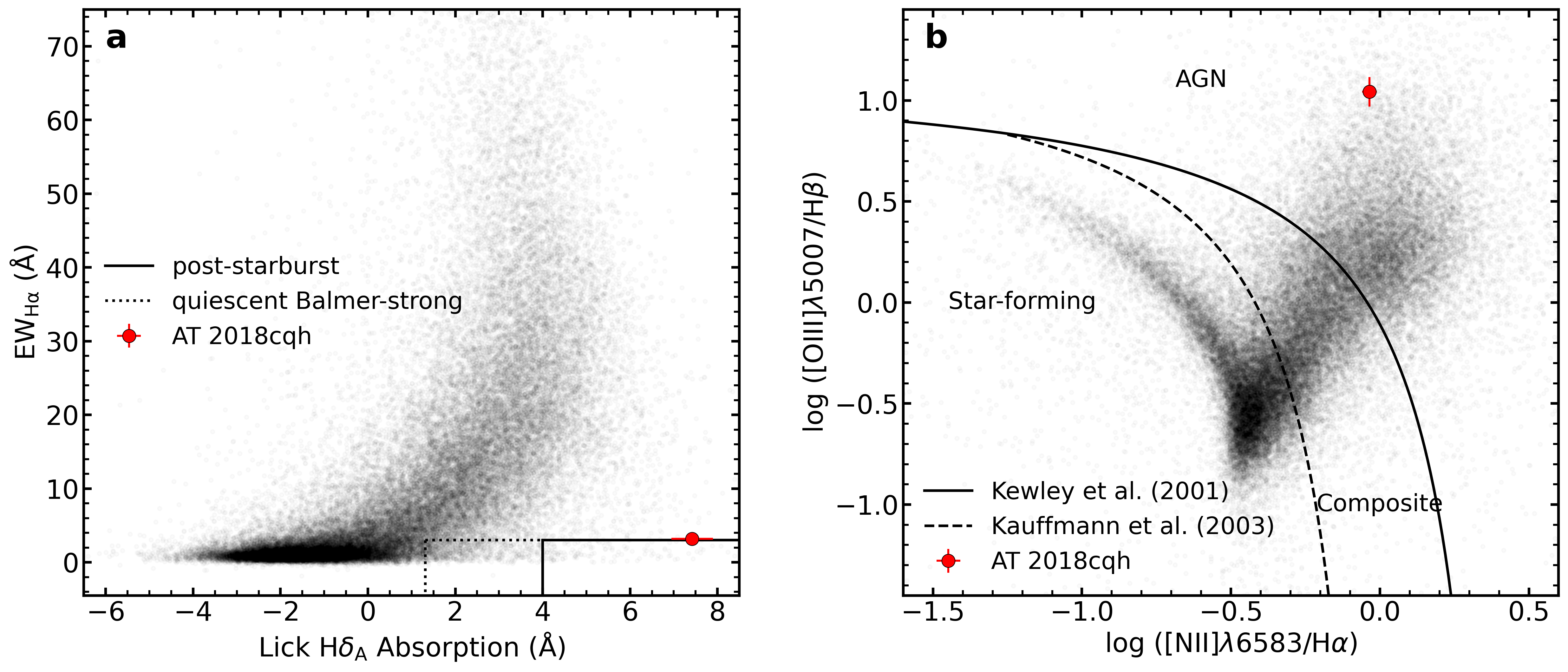}
\caption{\textbf{{\textbar} Spectral characteristics of the host galaxy of AT 2018cqh. a} the ${\rm H}\alpha$ equivalent width versus the Lick ${\rm H}\delta$ absorption index. The black solid line represents the division for post-starburst galaxies, while the dotted line indicates the boundary for quiescent Balmer-strong galaxies \cite{French+2016}. \textbf{b} the BPT diagnostic diagram, plotting the line ratios of [O \textsc{iii}]$\lambda$5007/${\rm H}\beta$ against [N \textsc{ii}]$\lambda$6584/${\rm H}\alpha$. The solid line is the demarcation between star-forming galaxies and narrow-line AGNs (Seyfert galaxies) \cite{Kauffmann+2003}. The dashed line shows the theoretical maximum for starburst galaxies \cite{Kewley+2001}. The background black dots correspond to typical SDSS galaxies \cite{French+2016}, shown for comparison. The error bars represent 1$\sigma$ uncertainties. Source data are provided as a Source Data file.}\label{fighost}
\end{figure}

\clearpage

\begin{figure}[p]
\addtocounter{Efigure}{1}
\centering
\includegraphics[width=0.9\textwidth]{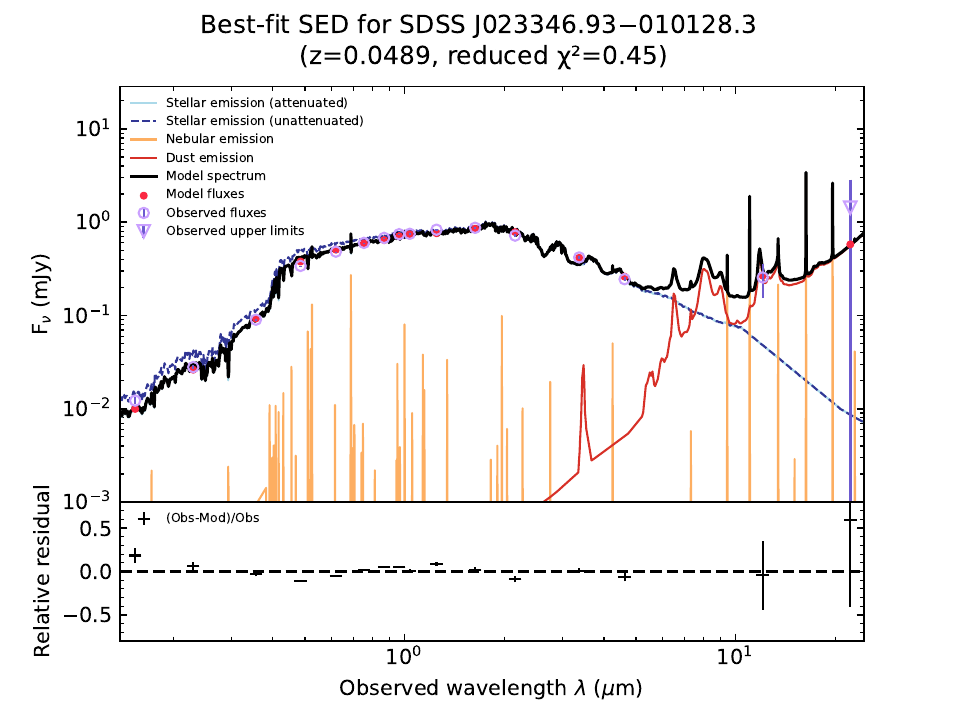}
\caption{\textbf{{\textbar} Best-fit pre-outburst SED for the host galaxy of AT 2018cqh.} The photometric data points from various bands are represented by purple circles for observed fluxes and purple triangles for upper limits. The black line corresponds to the best-fit model spectrum, while the red dots show the model fluxes at the same wavelengths as the observed data. The light blue line represents the attenuated stellar emission. The dark blue line shows the unattenuated stellar emission, reflecting the intrinsic stellar light. The orange line depicts the nebular emission. The red line represents the dust emission, modeling the thermal emission from dust in the galaxy. The error bars represent 1$\sigma$ uncertainties. Source data are provided as a Source Data file.}\label{figsed}
\end{figure}

\clearpage

\begin{table}[p]
\renewcommand\tablename{Supplementary Table}
\caption{\textbf{{\textbar} X-ray observational log of AT 2018cqh.}}\label{tabxlog}
\begin{tabular*}{\textwidth}{@{\extracolsep\fill}lccc}
\toprule%
Observation & Observation Time & Obsid & Exposure (s) \\
\midrule
Swift/XRT & & & \\
XRT1 & 2023-08-09T00:32:00 & 00016191001 & 4325.3 \\
XRT2\footnotemark[1] & 2024-06-14T05:56:00 & 00016191002 & 1698.1 \\
XRT3-1 & 2024-09-19T01:44:00 & 00016191003 & 2075.3 \\
XRT3-2 & 2024-09-22T10:13:00 & 00016191004 & 3381.4 \\
XRT3-3 & 2024-09-27T04:57:00 & 00016191005 & 979.0 \\
XRT3-4 & 2024-09-30T05:15:00 & 00016191006 & 1840.5 \\
XRT3-5 & 2024-10-02T00:05:00 & 00016191007 & 372.1 \\
XRT3-6 & 2024-10-06T19:01:00 & 00016191008 & 1585.8 \\
XRT3 (combined) & --- & --- & 10234.1 \\
XRT4-1 & 2024-10-21T18:48:00 & 00016191009 & 2447.3 \\
XRT4-2 & 2024-11-10T13:24:00 & 00016191010 & 611.8 \\
XRT4-3 & 2024-11-11T02:07:00 & 00016191011 & 222.3 \\
XRT4-4 & 2024-11-20T02:39:00 & 00016191013 & 1783.1 \\
XRT4-5 & 2024-11-30T10:52:00 & 00016191014 & 1366.0 \\
XRT4 (combined) & --- & --- & 6430.5 \\
XRT5-1 & 2024-12-09T09:44:00 & 00016191015 & 1925.4 \\
XRT5-2 & 2024-12-14T16:53:00 & 00016191016 & 2120.2 \\
XRT5-3 & 2024-12-19T03:59:00 & 00016191017 & 2120.2 \\
XRT5-4 & 2024-12-24T03:07:00 & 00016191018 & 1950.4 \\
XRT5-5 & 2024-12-29T06:46:00 & 00016191019 & 1111.3 \\
XRT5 (combined) & --- & --- & 9227.5 \\
\midrule
EP/FXT & & & \\
FXT1 & 2024-08-10T23:41:19 & 11904194932 & 11723 \\
\midrule
XMM/EPIC & & & \\
EPIC1 & 2025-01-22T17:42:20 & 0954191001 & 39000 \\
\bottomrule
\end{tabular*}
\footnotetext{Note: The Observation column identifies the instrument responsible for each observation and specifies which observations contributed to the data points shown in Fig. \ref{figlc}. The Observation Time column lists the UTC start time of each exposure. The Obsid column provides the unique observation ID for each observation. The Exposure Time column represents the total cleaned accumulated time, as individual exposures may not have been continuous. For XMM-Newton, the exposure time reflects the total exposure duration.}
\footnotetext[1]{XRT2 is excluded from spectral modeling because it did not achieve a 3$\sigma$ detection due to insufficient exposure time.}
\end{table}

\clearpage
\newgeometry{top=3cm, bottom=3cm, left=2cm, right=2cm}

\begin{table}[p]
\renewcommand\tablename{Supplementary Table}
\caption{\textbf{{\textbar} X-ray spectral fitting results.}}\label{tabxspec}
\begin{tabular}{@{}lccccccc@{}}
\toprule%
Observation & model & $\Gamma$ & ${kT}$ (eV) & ${T}_{\rm in}$ (eV) & C-stat/d.o.f & log Flux$_{\rm unabs}$ \\
\midrule
XRT1 & $powerlaw$ & ${3.91}_{-0.64}^{+0.77}$ & --- & --- & 27.24/43 & ${-12.47}_{-0.09}^{+0.09}$ \\
& $bbodyrad$ & --- & ${83.7}_{-19.2}^{+22.5}$ & --- & 36.20/43 & ${-12.52}_{-0.10}^{+0.10}$ \\
& $diskbb$ & --- & --- & ${119.0}_{-25.9}^{+23.8}$ & 34.58/43 & ${-12.51}_{-0.09}^{+0.09}$ \\
FXT1 & $powerlaw$ & ${3.68}_{-0.10}^{+0.10}$ & --- & --- & 245.35/275 & ${-12.45}_{-0.02}^{+0.02}$ \\
& $bbodyrad$ & --- & ${121.8}_{-4.3}^{+4.9}$ & --- & 470.73/275 & ${-12.54}_{-0.02}^{+0.02}$ \\
& $diskbb$ & --- & --- & ${177.4}_{-8.5}^{+9.8}$ & 405.70/275 & ${-12.54}_{-0.02}^{+0.02}$ \\
XRT3 & $powerlaw$ & ${3.73}_{-0.25}^{+0.26}$ & --- & --- & 68.48/290 & ${-12.43}_{-0.05}^{+0.05}$ \\
& $bbodyrad$ & --- & ${113.0}_{-9.4}^{+11.0}$ & --- & 100.29/290 & ${-12.50}_{-0.06}^{+0.05}$ \\
& $diskbb$ & --- & --- & ${163.1}_{-18.4}^{+19.3}$ & 92.92/290 & ${-12.50}_{-0.05}^{+0.05}$ \\
XRT4 & $powerlaw$ & ${4.45}_{-0.38}^{+0.41}$ & --- & --- & 31.24/410 & ${-12.40}_{-0.07}^{+0.07}$ \\
& $bbodyrad$ & --- & ${93.5}_{-11.5}^{+12.7}$ & --- & 38.53/410 & ${-12.47}_{-0.07}^{+0.07}$ \\
& $diskbb$ & --- & --- & ${125.2}_{-18.7}^{+18.6}$ & 36.69/410 & ${-12.46}_{-0.07}^{+0.07}$ \\
XRT5 & $powerlaw$ & ${3.36}_{-0.27}^{+0.27}$ & --- & --- & 45.75/366 & ${-12.47}_{-0.07}^{+0.06}$ \\
& $bbodyrad$ & --- & ${147.9}_{-18.0}^{+17.4}$ & --- & 70.92/366 & ${-12.57}_{-0.07}^{+0.06}$ \\
& $diskbb$ & --- & --- & ${210.9}_{-24.7}^{+27.9}$ & 62.71/366 & ${-12.54}_{-0.07}^{+0.06}$ \\
EPIC1 & $powerlaw$ & ${4.31}_{-0.04}^{+0.04}$ & --- & --- & 3802.20/7077 & ${-12.43}_{-0.01}^{+0.01}$ \\
& $bbodyrad$ & --- & ${76.6}_{-1.1}^{+1.0}$ & --- & 4748.69/7077 & ${-12.43}_{-0.01}^{+0.01}$ \\
& $diskbb$ & --- & --- & ${100.9}_{-1.8}^{+1.7}$ & 4572.52/7077 & ${-12.43}_{-0.01}^{+0.01}$ \\
\midrule
XRT1 & $simpl \times bbodyrad$ & ${1.67}_{-0.61}^{+0.64}$ & ${57.9}_{-12.0}^{+14.2}$ & --- & 17.85/43 & ${-12.42}_{-0.10}^{+0.10}$ \\
& $simpl \times diskbb$ & ${1.64}_{-0.62}^{+0.65}$ & --- & ${70.1}_{-16.1}^{+20.5}$ & 17.97/43 & ${-12.41}_{-0.10}^{+0.10}$ \\
FXT1 & $simpl \times bbodyrad$ & ${2.36}_{-0.24}^{+0.24}$ & ${63.5}_{-4.6}^{+4.6}$ & --- & 209.44/273 & ${-12.41}_{-0.03}^{+0.02}$ \\
& $simpl \times diskbb$ & ${2.32}_{-0.25}^{+0.25}$ & --- & ${78.0}_{-6.6}^{+6.7}$ & 210.24/273 & ${-12.41}_{-0.03}^{+0.02}$ \\
XRT3 & $simpl \times bbodyrad$ & ${2.03}_{-0.47}^{+0.56}$ & ${69.8}_{-10.8}^{+11.2}$ & --- & 60.53/290 & ${-12.41}_{-0.06}^{+0.06}$ \\
& $simpl \times diskbb$ & ${1.99}_{-0.49}^{+0.57}$ & --- & ${86.7}_{-15.5}^{+17.0}$ & 60.67/290 & ${-12.41}_{-0.06}^{+0.06}$ \\
XRT4 & $simpl \times bbodyrad$ & ${2.41}_{-1.09}^{+1.52}$ & ${70.2}_{-13.0}^{+13.6}$ & --- & 28.51/410 & ${-12.41}_{-0.06}^{+0.06}$ \\
& $simpl \times diskbb$ & ${2.29}_{-1.18}^{+1.63}$ & --- & ${87.5}_{-19.4}^{+21.3}$ & 28.79/410 & ${-12.40}_{-0.08}^{+0.07}$ \\
XRT5 & $simpl \times bbodyrad$ & ${2.34}_{-0.46}^{+0.47}$ & ${63.2}_{-13.8}^{+16.1}$ & --- & 40.90/366 & ${-12.44}_{-0.07}^{+0.07}$ \\
& $simpl \times diskbb$ & ${2.31}_{-0.47}^{+0.49}$ & --- & ${79.1}_{-19.9}^{+25.9}$ & 41.12/366 & ${-12.44}_{-0.07}^{+0.07}$ \\
EPIC1 & $simpl \times bbodyrad$ & ${2.60}_{-0.09}^{+0.09}$ & ${55.7}_{-1.1}^{+1.1}$ & --- & 3535.94/7075 & ${-12.40}_{-0.01}^{+0.01}$ \\
& $simpl \times diskbb$ & ${2.54}_{-0.10}^{+0.10}$ & --- & ${68.5}_{-1.5}^{+1.6}$ & 3544.50/7075 & ${-12.40}_{-0.01}^{+0.01}$ \\
\midrule
XRT1 & $zashift \times simpl \times bbodyrad$ & ${1.67}_{-0.61}^{+0.64}$ & ${60.7}_{-12.5}^{+14.9}$ & --- & 17.85/43 & ${-12.42}_{-0.10}^{+0.10}$ \\
& $zashift \times simpl \times bbodyrad$ & ${2.3}$ (fix) & ${56.5}_{-11.8}^{+14.6}$ & --- & 18.81/43 & ${-12.40}_{-0.10}^{+0.09}$ \\
& $zashift \times simpl \times diskbb$ & ${1.66}_{-0.62}^{+0.65}$ & --- & ${72.5}_{-16.8}^{+21.5}$ & 17.97/43 & ${-12.41}_{-0.10}^{+0.10}$ \\
& $zashift \times simpl \times diskbb$ & ${2.3}$ (fix) & --- & ${67.8}_{-16.2}^{+19.8}$ & 18.96/43 & ${-12.40}_{-0.10}^{+0.09}$ \\
FXT1 & $zashift \times simpl \times bbodyrad$ & ${2.36}_{-0.24}^{+0.24}$ & ${66.6}_{-4.9}^{+4.8}$ & --- & 209.44/273 & ${-12.41}_{-0.03}^{+0.02}$ \\
& $zashift \times simpl \times diskbb$ & ${2.32}_{-0.25}^{+0.25}$ & --- & ${81.8}_{-7.0}^{+7.0}$ & 210.24/273 & ${-12.41}_{-0.03}^{+0.02}$ \\
XRT3 & $zashift \times simpl \times bbodyrad$ & ${2.03}_{-0.47}^{+0.56}$ & ${73.3}_{-11.3}^{+11.8}$ & --- & 60.53/290 & ${-12.41}_{-0.06}^{+0.06}$ \\
& $zashift \times simpl \times bbodyrad$ & ${2.3}$ (fix) & ${70.6}_{-10.0}^{+10.9}$ & --- & 60.79/290 & ${-12.41}_{-0.06}^{+0.06}$ \\
& $zashift \times simpl \times diskbb$ & ${1.99}_{-0.49}^{+0.57}$ & --- & ${90.9}_{-16.3}^{+17.8}$ & 60.67/290 & ${-12.41}_{-0.06}^{+0.06}$ \\
& $zashift \times simpl \times diskbb$ & ${2.3}$ (fix) & --- & ${86.3}_{-14.1}^{+16.3}$ & 61.00/290 & ${-12.40}_{-0.06}^{+0.06}$ \\
XRT4 & $zashift \times simpl \times bbodyrad$ & ${2.41}_{-1.10}^{+1.52}$ & ${73.6}_{-13.7}^{+14.3}$ & --- & 28.51/410 & ${-12.41}_{-0.08}^{+0.08}$ \\
& $zashift \times simpl \times bbodyrad$ & ${2.3}$ (fix) & ${74.2}_{-11.1}^{+12.8}$ & --- & 28.51/410 & ${-12.41}_{-0.08}^{+0.07}$ \\
& $zashift \times simpl \times diskbb$ & ${2.29}_{-1.18}^{+1.63}$ & --- & ${91.8}_{-20.4}^{+22.3}$ & 28.79/410 & ${-12.40}_{-0.08}^{+0.07}$ \\
& $zashift \times simpl \times diskbb$ & ${2.3}$ (fix) & --- & ${91.7}_{-16.2}^{+19.5}$ & 28.79/410 & ${-12.40}_{-0.08}^{+0.07}$ \\
XRT5 & $zashift \times simpl \times bbodyrad$ & ${2.34}_{-0.46}^{+0.47}$ & ${66.2}_{-14.4}^{+16.9}$ & --- & 40.90/366 & ${-12.44}_{-0.07}^{+0.07}$ \\
& $zashift \times simpl \times bbodyrad$ & ${2.3}$ (fix) & ${66.7}_{-13.0}^{+15.8}$ & --- & 40.91/366 & ${-12.44}_{-0.07}^{+0.07}$ \\
& $zashift \times simpl \times diskbb$ & ${2.31}_{-0.47}^{+0.49}$ & --- & ${83.0}_{-20.9}^{+27.2}$ & 41.12/366 & ${-12.44}_{-0.07}^{+0.07}$ \\
& $zashift \times simpl \times diskbb$ & ${2.3}$ (fix) & --- & ${83.5}_{-19.8}^{+23.4}$ & 41.13/366 & ${-12.44}_{-0.07}^{+0.07}$ \\
EPIC1 & $zashift \times simpl \times bbodyrad$ & ${2.60}_{-0.09}^{+0.09}$ & ${58.4}_{-1.1}^{+1.1}$ & --- & 3535.90/7075 & ${-12.40}_{-0.01}^{+0.01}$ \\
& $zashift \times simpl \times diskbb$ & ${2.54}_{-0.10}^{+0.10}$ & --- & ${71.8}_{-1.6}^{+1.6}$ & 3544.42/7075 & ${-12.40}_{-0.01}^{+0.01}$ \\
\botrule
\end{tabular}
\footnotetext{Note: The spectral fitting utilized data from all Swift and EP observations within the energy range of 0.3--10 keV , and from all XMM-Newton observations within the range of 0.2--12 keV. The Observation column specifies the data points shown in Fig. \ref{figlc}. The model column indicates the fitted model corresponding to the fitting results for each entry. Each fit incorporates absorption from the Galactic column, modeled with $phabs$ (${N}_{\rm H} = 2.43 \times {10}^{20}$ cm$^{-2}$). $\Gamma$ denotes the photon index from the power-law model ($powerlaw$) or the empirical Comptonization model ($simpl$). $kT$ represents the temperature from the area-normalized blackbody model ($bbodyrad$). ${T}_{\rm in}$ indicates the apparent inner-disk temperature from the accretion disk model ($diskbb$). The C-stat/d.o.f column shows the goodness of fit. The unabsorbed flux of 0.3--2.0 keV band Flux$_{\rm unabs}$ is derived using the $cflux$ component. All quoted errors correspond to the 1$\sigma$ confidence level.}
\end{table}

\end{document}